\documentclass[12pt]{iopart}

\usepackage{graphicx}
\usepackage{bm}
\usepackage{iopams}

\begin{document}

\title{Universality of Long-Range Correlations in Expansion-Randomization Systems}

\author{P W Messer\dag, M L\"assig\ddag, P F Arndt\dag}

\address{\dag\ Max Planck Institute for Molecular Genetics, Ihnestr.~73, 14195 Berlin, Germany}
\address{\ddag\ Institute for Theoretical Physics, University of Cologne, Z\"ulpicher Str.~77, 50937 K\"oln, Germany}

\ead{philipp.messer@molgen.mpg.de}

\begin{abstract}
We study the stochastic dynamics of sequences evolving by 
single site mutations, segmental duplications, deletions, and random insertions. These processes are relevant for the 
evolution of genomic DNA. They define a universality 
class of non-equilibrium 1D expansion-randomization systems
with generic stationary long-range correlations in a regime
of growing sequence length. We obtain explicitly the two-point
correlation function of the sequence composition and the distribution function of the composition bias in sequences of finite length. The characteristic
exponent $\chi$ of these quantities 
is determined by the ratio of two effective rates,
which are explicitly calculated for several specific
sequence evolution dynamics of the universality class. 
Depending on the value of $\chi$, we find two different 
scaling regimes, which are distinguished by the detectability 
of the initial composition bias. All
analytic results are accurately verified by numerical
simulations. We also discuss the non-stationary build-up and decay of correlations, as well as more complex evolutionary scenarios, where the rates of the processes vary in time. Our findings provide a possible
example for the emergence of universality in molecular
biology. 
\end{abstract}



 
\maketitle

\section{Introduction}

Universality classes with long-range correlations are a
hallmark of systems with many degrees of freedom throughout
physics. In equilibrium condensed matter systems, they mark
critical points or phases with a particular symmetry. Out
of equilibrium, power-law correlations are more generic
but the classification of universality classes becomes more
difficult. Well known examples are surface growth, reaction-diffusion
systems, and self-organized criticality.

A striking example of long-range correlations in a
biological system has been found in the base pair
composition of genomic DNA more than a decade
ago~\cite{Li92,Peng92,Voss92}. Since then, the composition
correlations of DNA have been studied  extensively  by a
variety of different methods, and nowadays it is well
established that long-range correlations appear in the 
genomes of many
species~\cite{Arneodo95,Vieira99,Yu01,Bernaola02,Li04,Li05}.
The form of these correlations, however, is much more
complex than simple power-laws. Within one chromosome,
there is often a variety of different scaling regimes and
effective exponents, and sometimes no clear scaling at all.

Despite the ubiquity of long-range correlations in genomes,
little is known about their origin. A likely dynamical
scenario is that they are generated by the stochastic
processes of molecular sequence evolution. 
In~\cite{Messer05}, we have studied a minimal evolutionary 
dynamics producing long-range correlations that can be compared 
to DNA sequence data in a quantitative way. 
This dynamics incorporates {\em local} processes including 
single site mutations, duplications and
deletions of existing segments of the sequence, and
insertions of random segments. It is inspired by a model
introduced by Li in 1989~\cite{Li89,Li91}. We have proved
the emergence of long-range correlations
in this dynamics: the correlation function of the
generated sequences decays as $C(r)\propto r^{-\alpha}$
for large $r$, and we have obtained an exact
expression for the decay exponent $\alpha$. 

In the first part of this article (sections~\ref{definition}-\ref{finite_size_distribution}), we present a more detailed calculation of 
the expectation value of the two-point correlation function and the finite-size
distribution function of the sequence composition bias. We show
that these quantities exhibit consistent scaling and that their functional forms are given mathematically as 
solutions of simple differential equations. The resulting power-law behavior 
can be expressed in terms of a single basic exponent $\chi$, the scaling dimension
of the local composition bias. This exponent is determined 
by just two effective parameters, which are simple functions of the
rates of the elementary processes. As a function of $\chi$, we find two distinct 
scaling regimes. In the strong-correlation regime ($\chi < 1/2$), the ancestral
composition bias can be detected in arbitrarily long sequences, in the weak-correlation
regime ($\chi > 1/2)$, this is possible only up to a characteristic sequence length. 

In the second part of the article (sections~\ref{model_extensions} and~\ref{dynamical_correlations}), we analyze various generalizations of the
sequence evolution model introduced in~\cite{Messer05} and 
demonstrate that they form a consistent universality class 
of non-equilibrium processes with generic long-range
correlations. These processes are biased segmental insertions
as well as  mutations with biased rates, which break the
$Z_2$ symmetry of the original model. The purpose of this
generalization is two-fold. On the one hand, the extended
model is biologically more accurate, since there is strong
evidence for the presence of GC-content biased segmental
insertion processes~\cite{Lander01}, as well as biased
mutation rates~\cite{Arndt03} during evolutionary history. 
Taking into account these processes  proves crucial for practical data analysis.
On the other hand, the model conceptually delineates what are the essential
ingredients of this non-equilibrium universality class: Long-range 
correlations emerge from the
interplay of processes producing correlations
on short scales, exponential growth of sequence length, and 
local randomization processes. The universal scaling behavior is distinguished
from the symmetry breaking caused by biased mutation processes.  
Furthermore, we generalize the scaling picture to  dynamical aspects of the build-up and
decay of correlations in time. We conclude with a discussion of the role of 
universality in a biological context.

\section{Sequence evolution model}

\label{definition}

The stochastic evolution model generates sequences $S=(s_1, \dots, s_N)$ of
variable length $N(t)$. For simplicity, their letters are taken from a binary
alphabet; $s_k = \pm1$. The elementary evolutionary steps are single site
mutations, duplications and deletions of existing sequence segments of arbitrary
lengths, and insertion of random segments. In fact, these processes are assumed
to be the major local processes acting on genomic DNA sequences during
evolutionary history~\cite{Graur00}. Formally, the dynamics of the processes can
be defined by
\begin{eqnarray}
(\cdots,s,\cdots)&\quad\to\quad(\cdots,-s,\cdots)&\qquad\mbox{mutation
rate}\;\mu\nonumber\\
(\cdots,(s)_{\ell},\cdots)&\quad\to\quad(\cdots,(s)_{\ell},(s)_{\ell},
\cdots)&\qquad\mbox{duplication rate}\;\delta_{\ell}\nonumber\\
(\cdots,s,\cdots)&\quad\to\quad(\cdots,s,(x)_{\ell},\cdots)&\qquad\mbox{insertion rate}\;\gamma_{\ell}^+\nonumber\\
(\cdots,(s)_{\ell},\cdots)&\quad\to\quad(\cdots,\cdots)&\qquad\mbox{deletion
rate}\;\gamma_{\ell}^-,\label{model_definition}
\end{eqnarray}
where $(s)_{\ell}$ denotes an existing sequence
segment of length $\ell\geq 1$, and $(x)_{\ell}$ is a segment of
length $\ell$ with uniformly distributed random letters $x_i=\pm 1$. Note that
by convention we do not allow insertion of random segments prior to the first
sequence element. Duplication and insertion events introduce a new sequence
segment next to an existing one and shift all subsequent letters $\ell$
positions to the right, thereby increasing the sequence length by~$\ell$.
Conversely, deletions shorten the length by~$\ell$. We will restrict all
processes to a maximum range $\ell_{\rm max}$, i.e., all rates $\delta_{\ell}$,
$\gamma_{\ell}^+$, and $\gamma_{\ell}^-$ are zero for $\ell>\ell_{\rm max}$.
Repeatedly running the processes over a time $t$ produces a statistical ensemble
of sequences; the corresponding averages are denoted by $\langle \dots \rangle
(t)$. This ensemble is characterized by the rates of the processes and by the
initial sequence. 
If we focus on scales much larger than $\ell_{\rm max}$, the statistical
properties of the generated sequences will then turn out to be determined by
just two effective parameters, the asymptotic growth rate $\lambda$ and the
effective mutation rate $\mu_{\rm eff}$, defined by
\begin{eqnarray}
\lambda=\delta_{\rm eff}+\gamma^+_{\rm eff}-\gamma^-_{\rm eff}\label{lambda}\\
\mu_{\rm eff}=\mu+\frac{1}{2}\gamma^+_{\rm eff}\label{mu_effective}.
\end{eqnarray} 
Both are simple functions of the cumulative rates of the ``microscopic''
processes, 
\begin{equation}
\delta_{\rm eff}=\sum_{\ell=1}^{\ell_{\rm
max}}\ell\delta_{\ell},\quad\gamma_{\rm eff}^+=\sum_{\ell=1}^{\ell_{\rm
max}}\ell\gamma_{\ell}^+,\quad\mbox{and}\quad\gamma_{\rm
eff}^-=\sum_{\ell=1}^{\ell_{\rm max}}\ell\gamma_{\ell}^-.
\end{equation}
The implementation of a numerical simulation of this dynamics is described
in~\ref{numerical_implementation}. We use the simulations to verify analytically
derived results of the following sections.

\section{Sequence growth and average composition}

\label{growth_composition}

\subsection{Average sequence length}

Running the processes defined in~(\ref{model_definition}) on sequences will
change their lengths $N(t)$. The dynamics of $\langle N\rangle(t)$ averaged over
an ensemble of sequences is
\begin{equation}
\label{N_dynamics}
\frac{\partial}{\partial t}\langle N\rangle(t) =\left[\sum_{\ell=1}^{\langle
N\rangle(t)}\ell\sigma(\delta_{\ell}-\gamma_{\ell}^-)+\sum_{\ell=1}^{\ell_{\rm
max}}\ell\gamma_{\ell}^+\right]\langle N\rangle(t).
\end{equation}
The finite size correction factor $\sigma=1-(\ell-1)/\langle N\rangle(t)$
accounts for the fact that in a sequence of length $N(t)$ there are only
$N(t)-\ell+1$ possibilities to duplicate or delete a segment of length $\ell$.
Using the initial condition $N(t=0)=N_0$, the solution of~(\ref{N_dynamics}) in
the asymptotic regime, $\langle N\rangle (t)\gg \ell_{\rm max}$, is then given
by 
\begin{equation}
\label{average_length}
\langle N \rangle(t) = N_0\exp(\lambda t)
\end{equation}
with the asymptotic growth rate $\lambda$, as defined in~(\ref{lambda}).

\subsection{Average composition bias}

\label{average_bias}

The average composition of a sequence element $s_k$ is measured by the
expectation value $\langle s_k\rangle(t)$, and in the following we will show
that any initial bias decays due to mutations and random insertions. $\langle
s_k\rangle(t)$ can be written as the difference
\begin{equation}
\langle s_k\rangle(t)=P_k^+(t)-P_k^-(t),
\label{s_k_definition}
\end{equation}
where $P_k^+(t)$ and $P_k^-(t)$ denote the probabilities of finding $s_k=+1$ or
$s_k=-1$ at time~$t$. The Master equations for $P_1^{\pm}(t)$ of the first
sequence site $s_1$ are given by
\begin{equation}
\frac{\partial}{\partial t}
P_1^{\pm}(t)=\mu[P_1^{\mp}-P_1^{\pm}]+\sum_{\ell=1}^{\ell_{\rm
max}}\gamma^-_{\ell}\:[P_{\ell}^{\pm}-P_1^{\pm}].
\end{equation} 
Omitting deletion and starting with a single site $S(t=0)=(+1)$, we obtain 
\begin{equation}
\label{s_1_decay}
\langle s_1 \rangle (t)=\exp(-2\mu t).
\end{equation}
 If one additionally allows deletion, any initial bias of $s_1$ will even decay
faster.
  
\begin{figure} [t!]
\centering
\includegraphics[width=\linewidth]{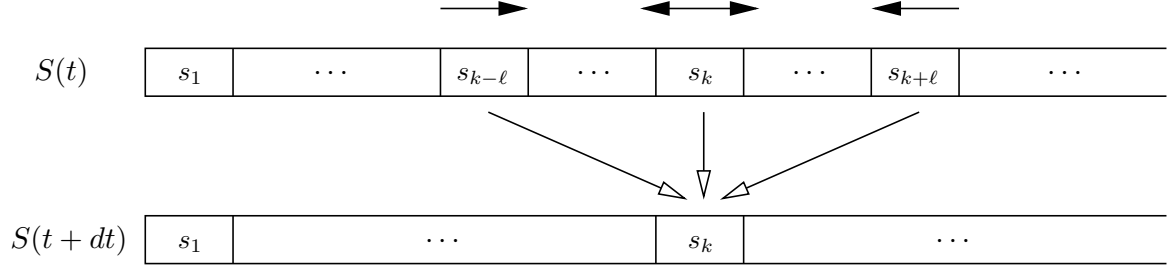}
\caption{\small Illustration of the different mechanisms contributing to
$\partial P^{\pm}_k(t)/\partial t$.\label{master_1point_eps}}
\end{figure}

Sequence sites $s_k$ at positions $k>1$ are also affected by duplications and
insertions, and the Master equations for the probabilities $P_k^{\pm}(t)$ take
the form 
\begin{equation}
\label{master_1point}
\eqalign{\fl\qquad\quad\frac{\partial}{\partial
t}P^{\pm}_k(t)=&\mu\:[P^{\mp}_k-P^{\pm}_k]+\sum_{\ell=1}^{\ell_{\rm max}}{\rm
min}(k-1,\ell)\:\gamma^+_{\ell}\:(1/2-P^{\pm}_k)\\
&+\sum_{\ell=1}^{k-2}(k-l-1)\:\gamma^+_{\ell}\:[P^{\pm}_{k-l}-P^{\pm}_k]+\sum_{
\ell=1}^{k-1}(k-l)\:\delta_{\ell}\:[P^{\pm}_{k-l}-P^{\pm}_k]\\
&+\sum_{\ell=1}^{\ell_{\rm max}}k\:\gamma^-_{\ell}\:[P^{\pm}_{k+l}-P^{\pm}_k].}
\end{equation}
The different mechanisms contributing to $\partial P_k^{\pm}(t)/\partial t$ are
illustrated in figure~\ref{master_1point_eps}. Any bias at site $s_k$ is again
diminished due to single-site mutations, as specified by the first term on the
r.h.s.~of~(\ref{master_1point}), but also by insertions of random segments
$(x_i,\dots,x_{i+\ell-1})$ of length $\ell$ at positions $i=k-\ell+1,\dots,k$,
which effectively randomize $s_k$ (second term). Additionally, there is a
``shift'' of composition bias from preceding sequence positions $s_{k-\ell}$ due
to insertions of random segments $(x_i,\dots,x_{i+\ell-1})$ of length $\ell$ at
positions $i=2,\dots,k-\ell$ (third term), or duplications of existing sequence
segments $(s_i,\dots,s_{i+\ell-1})$ with $i=1,\dots,k-\ell$ (fourth term).
Transport of bias from sites $s_{k+\ell}$ to $s_k$, on the other hand, occurs
due to deletion of existing segments $(s_i,\dots,s_{i+\ell-1})$ with
$i=1,\dots,k$ (last term).

\begin{figure} [t!]
\centering
\includegraphics[width=0.45\linewidth]{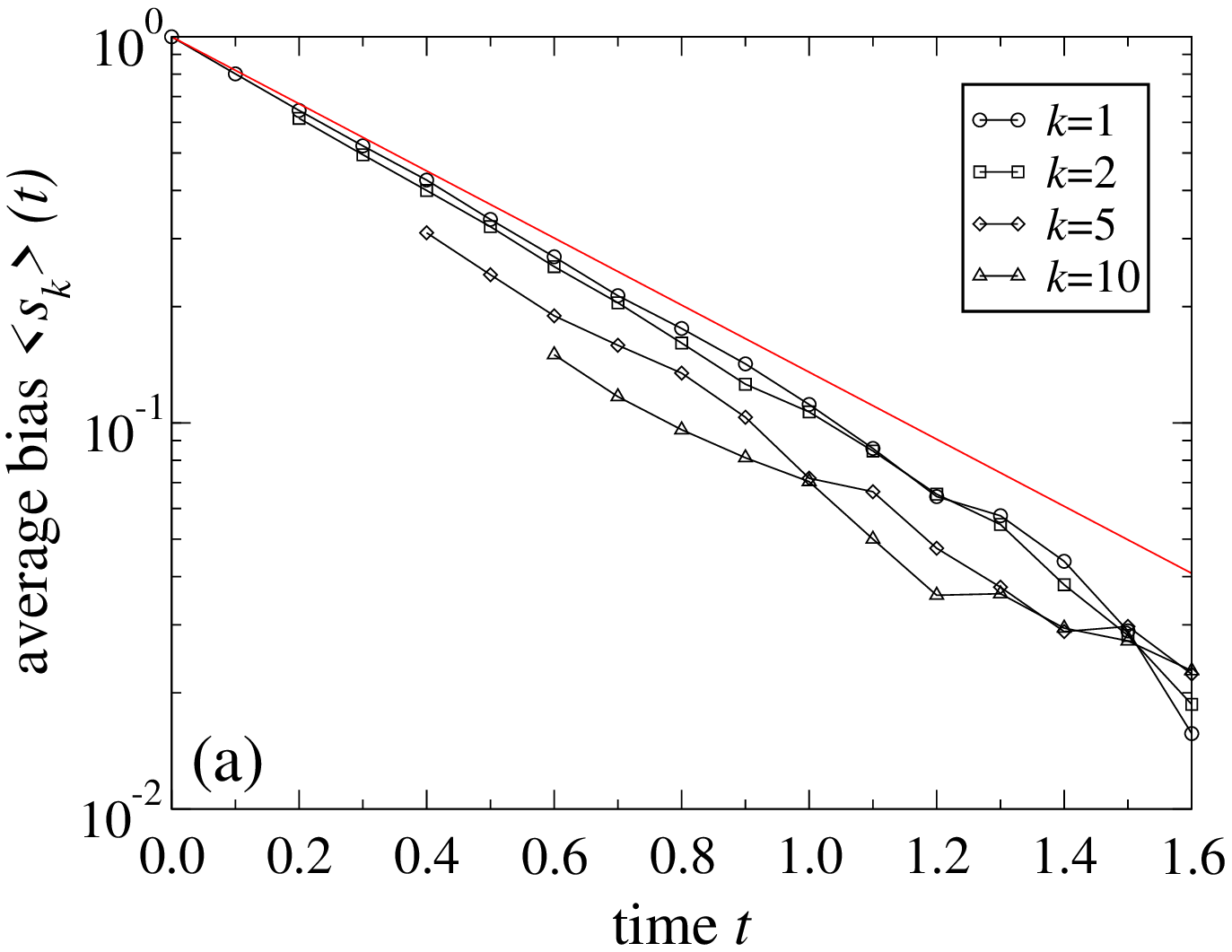}
\includegraphics[width=0.45\linewidth]{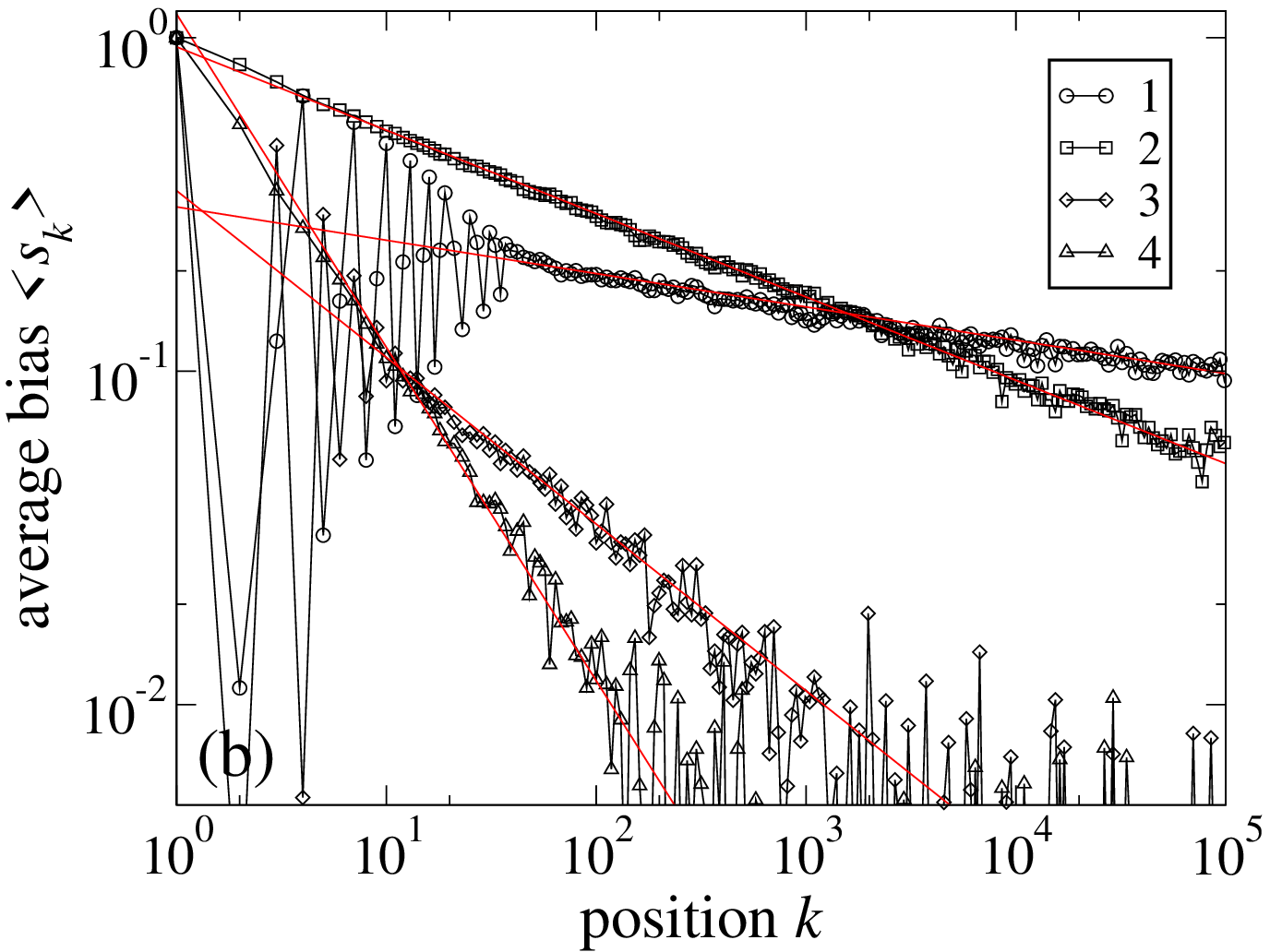}
\caption{\small Average composition bias $\langle s_k\rangle(t)$: (a) Decay of
$\langle s_k\rangle(t)$ in time for $k=1,2,5,10$. Rates of the processes are:
$\mu=1.0,\delta_1=4.0,\gamma^+_5=0.2,\gamma^-_2=0.5$. The red line is the
analytic lower bound on the rate of convergence~(\ref{s_k_decay}).(b) Stationary
$\langle s_k\rangle$ with fixed $\langle s_1\rangle=+1$ at different rates of the
elementary processes: (1) $\mu=1.0,\delta_3=15.0,\gamma^+_2=1.0,\gamma^-_7=1.0$;
(2) $\mu=1.0,\delta_1=16.0,\gamma^+_2=1.0,\gamma^-_1=2.0$; (3)
$\mu=1.0,\delta_2=6.0,\gamma^+_3=2.0,\gamma^-_4=0.5$; (4)
$\mu=1.0,\delta_1=4.0,\gamma^+_2=1.0,\gamma^-_4=0.5$. Red lines denote the
corresponding analytic asymptotics~(\ref{s_k_asymptotics}). All ensemble
averages were obtained by averaging over $10^6$ simulated sequences. 
\label{1point_eps}}
\end{figure}

In order to reveal the large-distance asymptotics of this dynamics for $k\gg
\ell_{\rm max}$ and in large sequences with $N(t)\gg \ell_{\rm max}$, we carry
out a continuum limit of~(\ref{master_1point}), i.e., we replace
the discrete index $k$ by a continuous variable and write 
$\langle s(k,t)\rangle\equiv\langle s_k\rangle(t)$ .  Using~(\ref{s_k_definition}) we
obtain a differential equation describing the asymptotic dynamics,  
\begin{equation}
\label{s_k_asymptotics_dgl}
\frac{\partial}{\partial t} \langle s(k,t) \rangle =-2\mu_{\rm eff}
\langle s(k,t) \rangle -
\lambda k\frac{\partial}{\partial k} \langle s(k,t) \rangle,
\end{equation}
with the asymptotic growth rate $\lambda$ and the effective mutation rate
$\mu_{\rm eff}$ defined in~(\ref{lambda}) and~(\ref{mu_effective}). The
transport of composition bias due to the net exponential expansion of the
sequences thereby gets incorporated in a dilatation operator of the functional
form $k\partial/\partial k$; all finite size effects vanish in this regime. 
Equation~(\ref{s_k_asymptotics_dgl}) 
has a solution of the form
\begin{equation}
\langle s(k,t) \rangle = e^{-2\mu_{\rm eff} t} \mathcal{S}(ke^{-\lambda t}),
\label{sol}
\end{equation}
where $\mathcal{S}(x)$ is a scaling function. This solution describes two different regimes of the expectation value,
depending on the boundary condition chosen. (a) With fixed initial condition
$s_1(t=0) = 1$, we have for any fixed $k$
\begin{equation}
\label{s_k_decay}
\langle s(k,t)\rangle \propto \exp(-2\mu_{\rm eff} t).
\end{equation}
as shown in Figure~\ref{1point_eps}(a) 
for different values of $k$ and a given set
of process rates. Thus, 
$\langle s(k,t)\rangle=0$ for all $k$ in the limit $t\to\infty$. (b) With fixed 
boundary condition $\langle
s_1\rangle=+1$ for all $t$ (i.e., suppressing mutations of the first element), 
we obtain a power-law decay of the composition bias
along the sequence,
\begin{equation}
\label{s_k_asymptotics}
\langle s(k)\rangle \propto k^{-\chi} \;\;\; 
\mbox{with }\chi = {\frac{2\mu_{\rm eff}}{\lambda}}. 
\end{equation} 
Numerical verification of the asymptotics~(\ref{s_k_asymptotics}) for this type
of dynamics is presented in Figure~\ref{1point_eps}(b), where we show the
measured $\langle s_k\rangle$ in ensembles of sequences with different sets of
rates using the simulation algorithm described
in~\ref{numerical_implementation}.

\section{Stationary two-point correlations}

\label{stationary_correlations}

\subsection{Master equation}

\begin{figure} [t!]
\centering
\includegraphics[width=\linewidth]{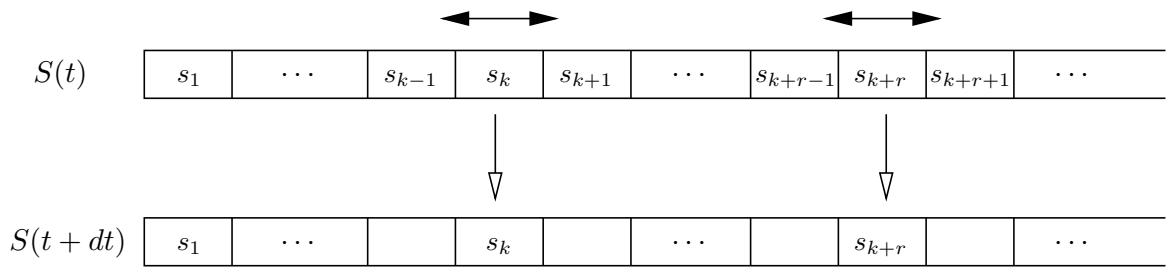}
\caption{\small Illustration of the different mechanisms contributing to the
dynamics of $P_{\rm eq}(k,r,t)$. Effectively mutational events are those that
randomize either $s_k$, or $s_{k+r}$. ``Expansion'' or ``contraction'' transport
of joint probability from $P_{\rm eq}(k,r\pm 1)$ to $P_{\rm eq}(k,r)$ occurs due
to duplication, insertion, or deletion events at sequence positions between
$s_k$ and $s_{k+r}$. ``Horizontal'' shift from $P_{\rm eq}(k\pm 1,r)$ to $P_{\rm
eq}(k,r)$ takes place if a duplication, insertion, or deletion occurs at
sequence positions prior to $s_k$.\label{master_2point_eps}}
\end{figure}

The dynamics of the composition correlation function $C(k,r,t) = \langle s_k
s_{k+r} \rangle (t)$ between two sequence positions $s_k$ and $s_{k+r}$ can be
derived by writing it as 
\begin{equation}
\label{C_definition}
C(k,r,t)=P_{\rm eq}(k,r,t)-P_{\rm op}(k,r,t),
\end{equation}
where $P_{\rm eq/op}(k,r,t)$ denote the joint probabilities of simultaneously
finding two equal or opposite symbols, respectively, at sequence positions $k$
and $k+r$ and time $t$. For simplicity, we start with a restricted sequence
evolution model where all processes are limited to single sequence sites
($\ell_{\rm max}=1$). The Master equation for $P_{\rm eq}(k,r,t)$ in the
single-site model takes the form
\numparts
\begin{eqnarray} 
\fl\qquad\quad\frac{\partial}{\partial t}P_{\rm eq}(k,r,t)=&2\mu\:[P_{\rm
op}(k,r)-P_{\rm eq}(k,r)]\label{master_a}\\
&+1/2\:\gamma^+_1\:[P_{\rm op}(k,r)-P_{\rm eq}(k,r)]\label{master_b}\\
&+1/2\:\gamma^+_1\:[P_{\rm op}(k-1,r)-P_{\rm eq}(k-1,r)]\label{master_c}\\
&+1/2\:\gamma^+_1\:[P_{\rm eq}(k-1,r)-P_{\rm eq}(k,r)]\label{master_d}\\
&+[(r-1)\gamma^+_1+r\delta_1]\:[P_{\rm eq}(k,r-1)-P_{\rm
eq}(k,r)]\label{master_e}\\
&+r\gamma^-_1\:[P_{\rm eq}(k,r+1)-P_{\rm eq}(k,r)]\label{master_f}\\
&+[(k-2)\gamma^+_1+(k-1)\delta_1]\:[P_{\rm eq}(k-1,r)-P_{\rm
eq}(k,r)]\label{master_g}\\
&+k\gamma^-_1\:[P_{\rm eq}(k+1,r)-P_{\rm eq}(k,r)]\label{master_h}.
\end{eqnarray}
\endnumparts
The different mechanisms contributing to $\partial P_{\rm eq}(k,r,t)/\partial t$
are illustrated in figure~\ref{master_2point_eps} and will now be discussed in
order. (\ref{master_a}) describes the change in $P_{\rm eq}(k,r,t)$ due to
mutation of any of the two sites (therefore two possibilities) in a pair of
equal or opposite symbols at positions $k$ and $k+r$. (\ref{master_b}) treats
the insertion of a random site at position $k+r$, which in half of the cases
will switch a pair of equal symbols $s_k=s_{k+r}$ to opposing symbols
$s_k=-s_{k+r}$, while two opposing symbols might be switched to equal symbols,
accordingly. A similar contribution arises from a random insertion at position
$k$. However, such an event can be regarded as duplication of $s_{k-1}$ with a
successional mutation of the newly introduced element $s_k$ in half of the cases.
If such a mutation occurs, the event is equivalent to~(\ref{master_b}) with the
difference that contributions of this processes to $\partial P_{\rm
eq}(k,r,t)/\partial t$ do now depend on the joint probabilities $P_{\rm
eq/op}(k-1,r,t)$ (\ref{master_c}). In the other half of the cases, where the
newly inserted random element $s_k$ is equal to $s_{k-1}$, the process causes a
shift of joint probability from $P_{\rm eq}(k-1,r,t)$ to $P_{\rm
eq}(k,r,t)$~(\ref{master_d}). Transport of joint probability at distance $r-1$
to such at distance $r$ takes place if a random site is inserted at sequence
positions $k+1,\dots,k+r-1$, or if any site $s_k,\dots,s_{k+r-1}$ is
duplicated~(\ref{master_e}). On the other hand, deletion of any
$s_{k+1},\dots,s_{k+r}$ produces a transport of joint probability from distance
$r+1$ to $r$~(\ref{master_f}). Despite this ``expansion'' and ``contraction''
transport of joint probability from distances $r+1$ or $r-1$ to $r$ at fixed
$k$, there is also a ``horizontal'' shift along the sequence: insertion of a
random site at positions $2,\dots,k-1$ or duplication of any site
$s_1,\dots,s_{k-1}$ shifts joint probability $P_{\rm eq}(k-1,r,t)$ to $P_{\rm
eq}(k,r,t)$~(\ref{master_g}), while deletion of an $s_1,\dots,s_k$ shifts
$P_{\rm eq}(k+1,r,t)$ to $P_{\rm eq}(k,r,t)$~(\ref{master_h}). 

Since we are interested in a stationary solution of this dynamics, we have to
consider the limit $t\to\infty$. It has already been shown in section~\ref{average_bias} that asymptotically $\langle s_k\rangle(t)\to 0$ for
large $t$ at all $k$. Furthermore, all processes are acting homogeneously along
the sequence, and therefore we expect the joint probabilities also to be
independent of $k$ in the long-time limit, i.e., $P_{\rm eq/op}(k,r)=P_{\rm
eq/op}(k\pm1,r)$ (verification is given by our numerical simulations). The
dynamics~(18) then simplifies to
\begin{eqnarray} 
\fl\qquad\quad\frac{\partial}{\partial t}P_{\rm
eq}(r,t)=&(2\mu+\gamma_1^+)\:[P_{\rm op}(r)-P_{\rm eq}(r)]\nonumber\\
&+[(r-1)\gamma^+_1+r\delta_1]\:[P_{\rm eq}(r-1)-P_{\rm eq}(r)]\\
&+r\gamma^-_1\:[P_{\rm eq}(r+1)-P_{\rm eq}(r)]\nonumber.
\end{eqnarray}
By exchanging $P_{\rm eq}$ and $P_{\rm op}$, we can state an equivalent equation
for $P_{\rm op}(r,t)$. Using~(\ref{C_definition}), we obtain the dynamics of the
correlation function $C(r,t)$ for large $t$
\begin{eqnarray}
\label{master_equation_C_single}
\fl\qquad\quad\frac{\partial}{\partial
t}C(r,t)=&-(4\mu+2\gamma_1^+)\:C(r)\nonumber\\
&+[(r-1)\gamma^+_1+r\delta_1]\:[C(r-1)-C(r)]\\
&+r\gamma^-_1\:[C(r+1)-C(r)]\nonumber.
\end{eqnarray}
This equation for the dynamics of $C(r,t)$ in the single-letter model
$(\ell_{\rm max}=1)$ is valid for all distances $r$ in the limit $t\to\infty$. A
corresponding dynamics can, in principle, be obtained analogously for the
general model with $\ell_{\rm max}>1$, although it will be more complicated due
to finite size effects coming into play for $r<\ell_{\rm max}$. However, for
large distances $r\gg\ell_{\rm max}$, these finite size effects can be
neglected, and the asymptotic dynamics of $C(r,t)$ in the general segmental
model is then given by 
\begin{eqnarray}
\label{master_equation_C}
\fl\qquad\quad\frac{\partial}{\partial t}C(r,t)=&-4\mu_{\rm
eff}\:C(r)\nonumber\\
&+\sum\nolimits_{\ell=1}^{\ell_{\rm
max}}\:[(r-\ell)\gamma_{\ell}^++(r-\ell+1)\delta_{\ell}]\:[C(r-\ell)-C(r)]\\
&+\sum\nolimits_{\ell=1}^{\ell_{\rm
max}}\:r\gamma_{\ell}^-\:[C(r+\ell)-C(r)]\nonumber
\end{eqnarray}
with the effective mutation rate $\mu_{\rm eff}$, as defined
in~(\ref{mu_effective}). Note that the dynamics~(\ref{master_equation_C_single})
of the single-letter model is a special case of the general
dynamics~(\ref{master_equation_C}) with $\ell_{\rm max}=1$.  

\subsection{Stationary solutions}

\label{stationary_solutions}

In the following, we will derive analytic solutions of the stationary
correlations $C(r)$ in our model. We start with the special case of only
single-site duplications and mutations ($\mu,\delta_1> 0$, all other rates are
zero). In this case, the solution of the dynamics~(\ref{master_equation_C}) in
the stationary state, $\partial C(r,t)/\partial t=0$, obeys the recursion
equation
\begin{equation} 
\label{C_dup_mut_recursion}
C(r)=\frac{r}{\alpha+r}\:C(r-1)\quad\mbox{with}
\quad\alpha=\frac{4\mu}{\delta_1}.
\end{equation} 
Using $C(0)\equiv 1$, the recursion can easily be solved,
yielding
\begin{equation}
C(r)=\prod_{n=1}^r\frac{n}{\alpha+n}.
\end{equation}
Introducing the gamma function and the beta function, defined by
\begin{equation}
\Gamma(x)=\int_0^{\infty}\:e^{-t}t^{x-1}dt,\quad\quad
B(x,y)=\frac{\Gamma(x)\Gamma(y)}{\Gamma(x+y)},
\end{equation}
$C(r)$ can finally be rewritten in the form
\begin{equation} 
\label{C_dup_mut_result}
C(r)=\frac{\Gamma(r+1)\Gamma(1+\alpha)}{\Gamma(r+1+\alpha)}=\alpha B(r+1,\alpha).
\end{equation}
To investigate the asymptotic regime, we evaluate the asymptotic behavior of
$B(r,\alpha)$ for $r\gg1$ which, in general, is given by
\begin{equation}
B(r,\alpha)\propto\Gamma(\alpha)\:r^{-\alpha}
\left[1-\frac{\alpha(\alpha-1)}{2r}\left(1+O
\left(\frac{1}{r}\right)\right)\right].
\end{equation} 
Applying this asymptotics to equation~(\ref{C_dup_mut_result}) we obtain
\begin{equation} 
\label{csym}
C(r)\propto r^{-\alpha}.
\end{equation}
Hence, we have proven the existence of long-range correlations in the simplified
single-site duplication-mutation model. The exponent $\alpha$ is determined by a
simple balance between the randomization processes (mutations) and the expansion
processes (duplications) which create correlations between neighboring sites and
transport these correlations to larger distances due to an overall expansion of
the system.

\begin{figure} [t!]
\centering
\includegraphics[width=0.45\linewidth]{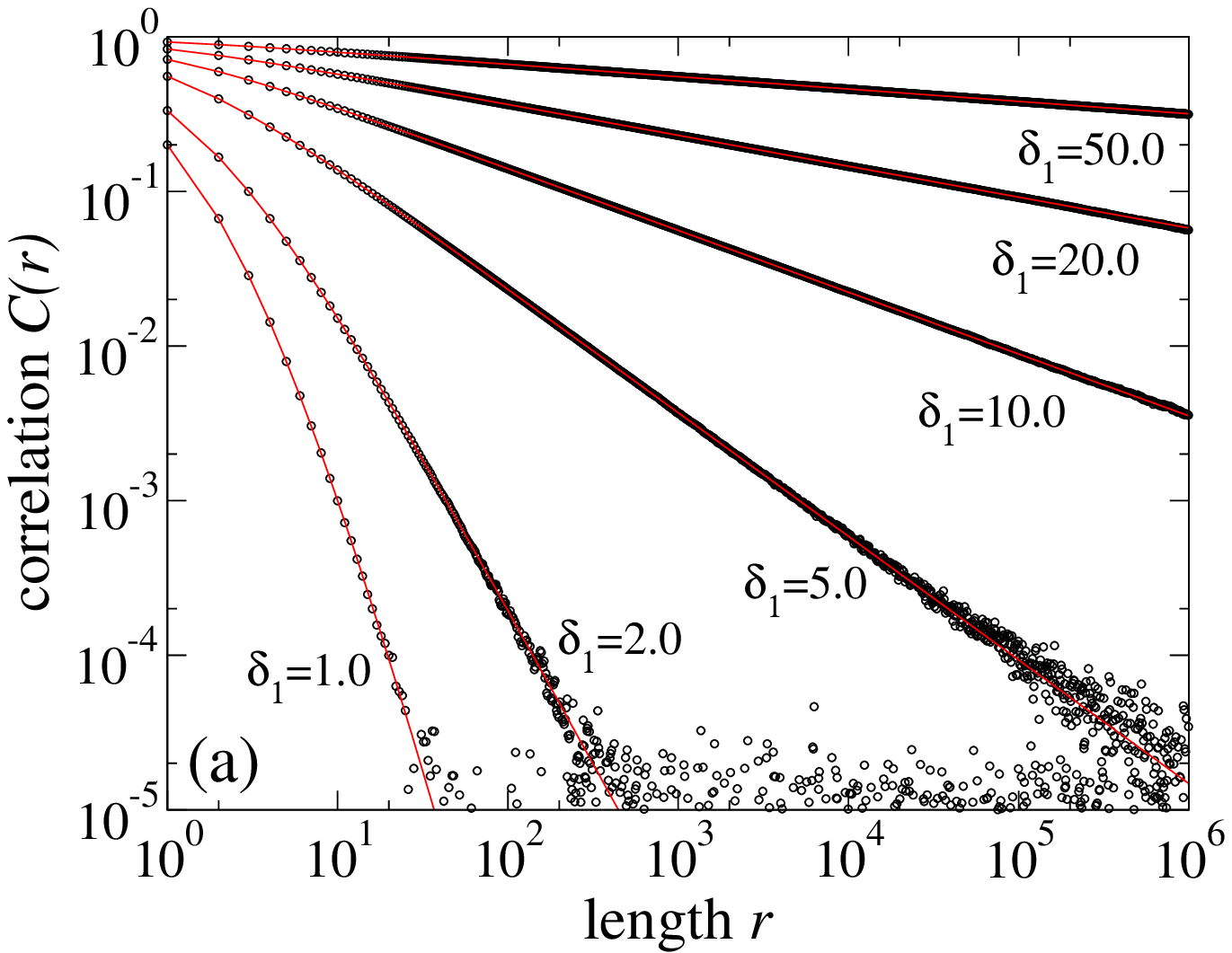}
\includegraphics[width=0.45\linewidth]{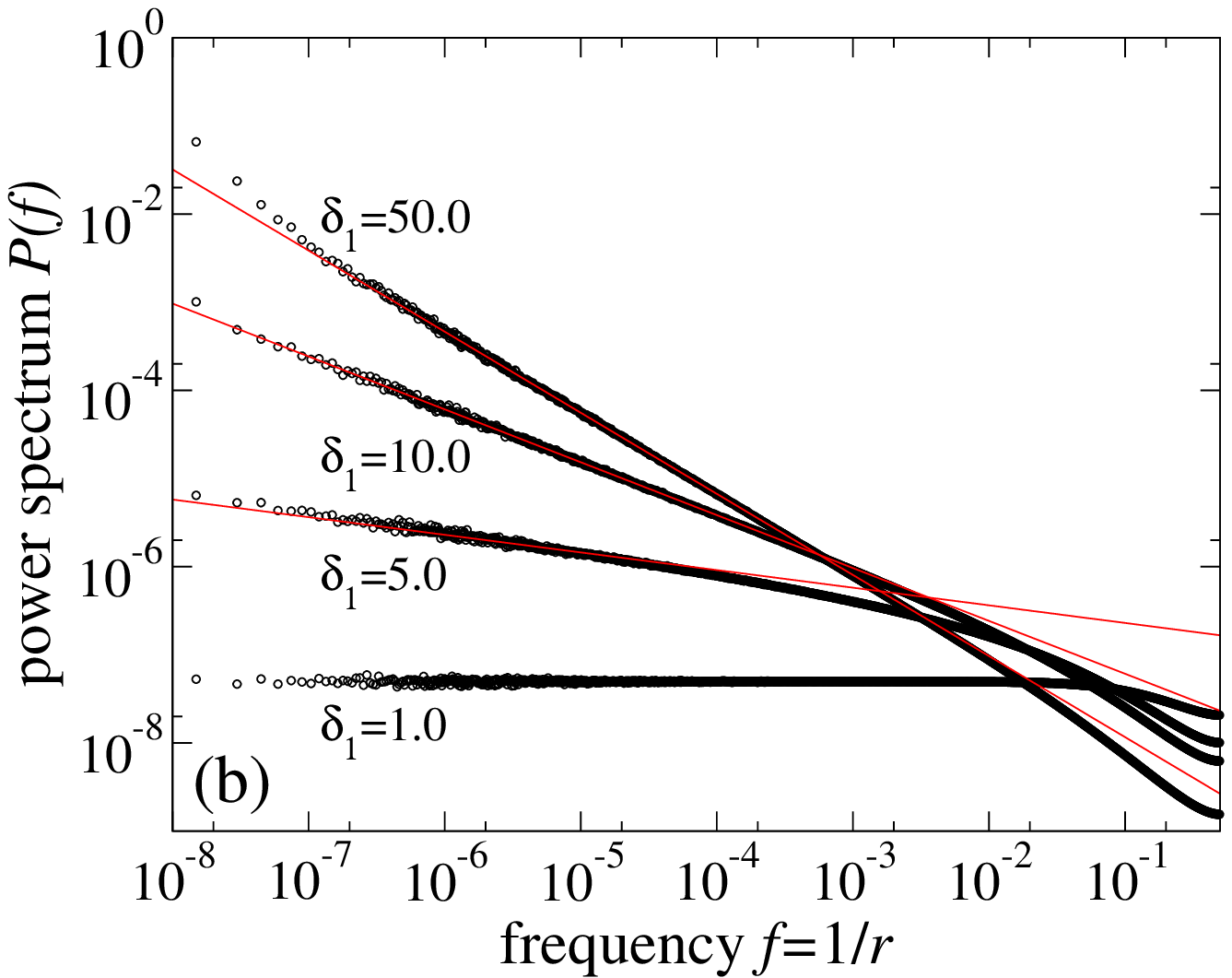}
\caption{\small Single-site duplication-mutation model: (a) Stationary
composition correlation $C(r)$ at different rates of the elementary processes;
numerical results (circles) and the analytic form~(\ref{C_dup_mut_result})
(lines) for $\mu=1.0$, $\delta_1$ varying. $C(r)$ is averaged along the
sequence. (b) Power spectra of simulated sequences for $\mu=1.0$ and $\delta_1$
varying: numerical results (circles) with the analytically predicted
$P(f)\propto f^{-\beta}$ in those cases where $\delta_1\ge5$ (lines). The
dynamics of the sequences was simulated until they reached a length of
$N=2^{27}\approx10^8$. All data sets were obtained by averaging over 100 runs. 
\label{dup_mut_numerics_plot}}
\end{figure}

We have performed extensive Monte Carlo simulations of this model using the
algorithm presented in~\ref{numerical_implementation}.
Figure~\ref{dup_mut_numerics_plot}(a) shows the numerical $C(r)$ for the
duplication-mutation dynamics with various rates of $\delta_1$ and $\mu$, which
is in excellent agreement with the analytic expression~(\ref{C_dup_mut_result}).

For reasons of comparability with former studies~\cite{Li89,Li91}, we also
calculated power spectra of the simulated sequences. In the stationary state,
the power spectrum $P(f)$ is the Fourier transform of the correlation function
$C(r)$. In our case, the large distance asymptotics of the correlation function
is given by $C(r)\propto r^{-\alpha}$, and the power spectrum will therefore
also decay algebraically, i.e., $P(f)\propto f^{-\beta}$ with the exponent
$\beta=1-\alpha$~\cite{Stanley94}. The resulting data is shown in
figure~\ref{dup_mut_numerics_plot}(b). Due to the fact that $C(r)\propto
r^{-\alpha}$ does only hold in the limit of $r\gg 1$, the analytically estimated
scaling $P(f)\propto f^{-\beta}$ is present at lower frequencies, but crosses
over to a different behavior at higher ones. At values of $\alpha\:>\:1$, $C(r)$
decays below the fluctuation threshold $\Delta C=1/\sqrt{N(t)}$~\cite{Weiss98},
before the scaling gets established, thus obviating the appearance of positive
exponents $\beta$. In those cases, we measure a flat power spectrum in the low
frequency part as one expects for a random sequence. The finite size deviations
of $C(r)$ at very large $r$ show up in the very low frequency part of the power
spectra, too. 

\begin{figure} [t!]
\centering
\includegraphics[width=0.75\linewidth]{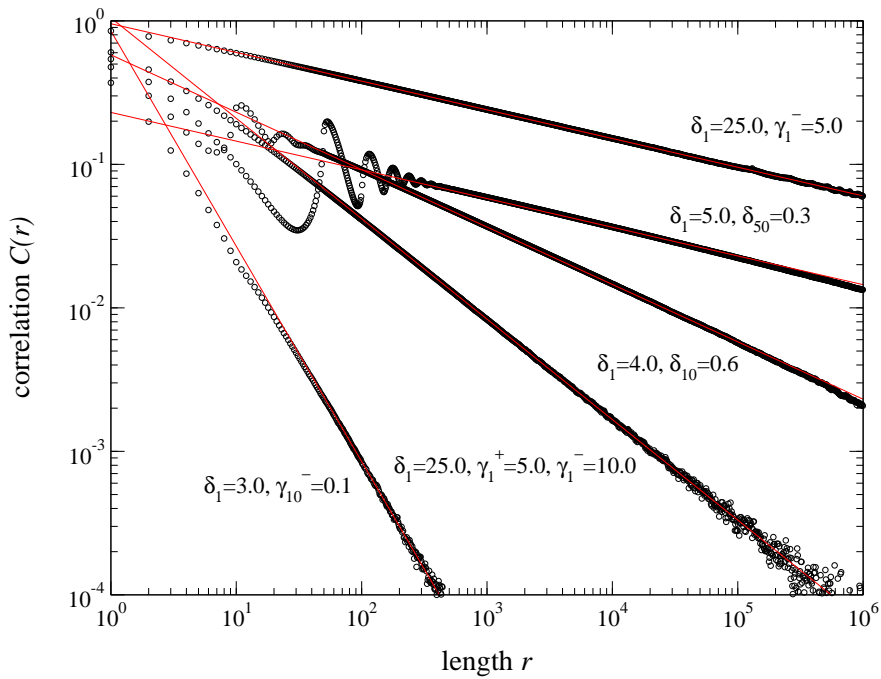}
\caption{\small Stationary $C(r)$ at different rates of the elementary processes
for the general model with various segmental processes present: Numerical
results (circles) with the analytic asymptotics~(\ref{C_asymptotics}) (lines)
for $\mu=1.0$ and varying rates of the other processes (rates not specified in
the plot are zero).\label{regions_plot}}
\end{figure} 

Obviously, one cannot expect the stationary $C(r)$ of the general model to be
described by a similar simple expression as has been obtained for the
single-site duplication-mutation dynamics in~(\ref{C_dup_mut_result}). Consider,
for example, a segmental duplication process, copying
segments of length $\ell_1=50$. In case this is the only duplication process present, it will introduce a peak in $C(r)$ at a distance corresponding to its segment
length $r=\ell_1$. If there is an additional duplication processes present, e.g.~one with $\ell_2=1$, the peak in $C(r)$ established by the first duplication process will be shifted to larger distances by the second process. The functional form of $C(r)$ will thus show complex behavior on short scales reflecting the ``microscopic'' details of the
elementary processes (see figure~\ref{regions_plot}). But what about the
large-distance asymptotics of $C(r)$ for $r\gg \ell_{\rm max}$ ? In this regime,
the dynamics of $C(r,t)$ is given by equation~(\ref{master_equation_C}).
Carrying out a continuum limit, the difference
equation~(\ref{master_equation_C}) can again be written as a simple differential
equation,
\begin{equation}
\label{C_continuum}
\frac{\partial}{\partial t}C(r,t)=-4\mu_{\rm eff}C(r,t)-\lambda
r\frac{\partial}{\partial r}C(r,t).
\end{equation} 
The stationary solution of equation~(\ref{C_continuum}) immediately yields the
power-law decay 
\begin{equation}
\label{C_asymptotics}
C(r)\propto r^{-\alpha}\quad\mbox{with}\quad\alpha=2\chi=\frac{4\mu_{\rm
eff}}{\lambda}. 
\end{equation}
Hence, on macroscopic distances $r\gg \ell_{\rm max}$ our model universally
produces long-range correlations in the sequences, irrespectively of the microscopic
details of the individual processes. The decay exponent $\alpha$ depends on only
two effective parameters which are simple functions of the rates of the
processes. Using these analytic results, we furthermore can qualitatively
classify the four different types of processes whether they increase $\alpha$,
or decrease it. Duplications are the only processes with $\partial
\alpha/\partial \delta_{\ell}<0$, since they raise the growth rate $\lambda$,
but have no effectively mutational influence on large scales. All other
processes, in contrast, will lead to larger values of $\alpha$ and thus to
faster decaying correlations along the sequence by an increase of their rates.

To verify these analytic results, we show the measured correlation functions
$C(r)$ of simulated sequences with all sorts of different processes present in
figure~\ref{regions_plot}. While on short scales the correlations reveal the
microscopic details of the particular processes, in the asymptotic regime
long-range correlations are ubiquitous. Their functional form is accurately
described by our analytics~(\ref{C_asymptotics}) with the effective
rates~(\ref{lambda}) and~(\ref{mu_effective}).

\section{Finite-size distribution of the composition bias}

\label{finite_size_distribution}

Up to this point, we have discussed correlation functions, which are defined as averages
over an ensemble of sequences generated by the same stochastic dynamics. What can we
say about the data of a single sequence, i.e., a single realization of the stochastic process?
To address this question, 
we now consider the distribution of the composition bias evaluated in finite sequence intervals $k,\dots,k+L-1$ of length $L$, 
\begin{equation}
\label{m}
m =\frac{1}{L}\sum_{k'=k}^{k+L-1} s_{k'}. 
\end{equation}
Generalizing equations~(\ref{s_k_asymptotics_dgl}) and~(\ref{C_continuum}), we obtain the following differential equation for the distribution function $P(m,L,t)$,
\begin{eqnarray}
\label{dynamics_Pm}
\frac{\partial}{\partial t}P(m,L,t) & = & -\lambda L\frac{\partial}{\partial L}P(m,L,t)
\nonumber \\
& & 
+2\mu_{\rm eff}\frac{\partial}{\partial m}[mP(m,L,t)]
+\frac{2\mu_{\rm eff}}{L} \frac{\partial^2}{\partial m^2} P(m,L,t),
\end{eqnarray}
which is valid again in a continuum approximation for $L \gg 1$. 
The three terms on the r.h.s.~describe, in order, the transport of the composition
bias due to the exponential dilatation of the sequence, its dissipative decay, and its stochastic
fluctuations. Notice that the last two terms are caused by the same basic mutation process and 
are therefore both proportional to $\mu_{\rm eff}$.   

\begin{figure} [t!]
\centering
\includegraphics[width=0.45\linewidth]{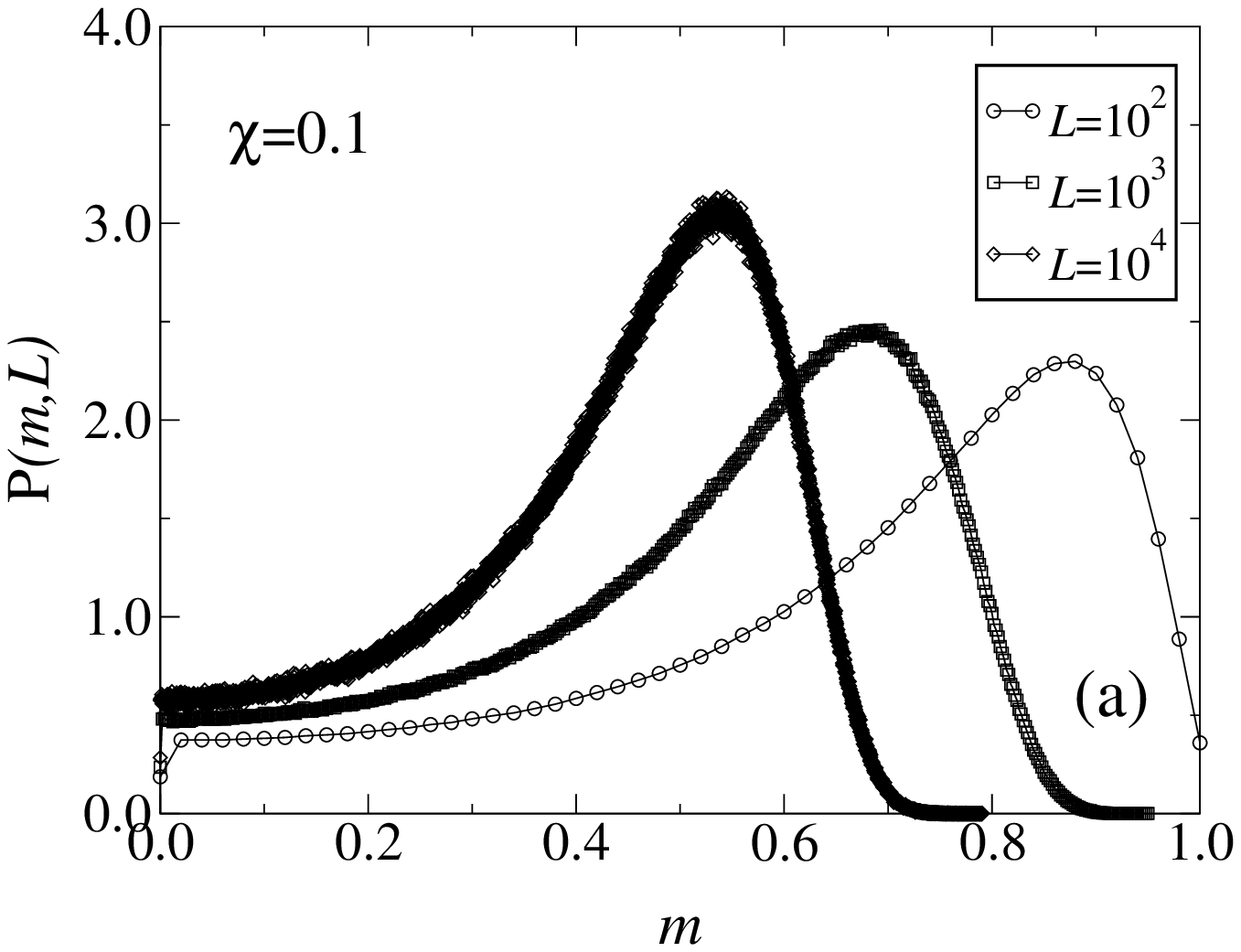}
\includegraphics[width=0.45\linewidth]{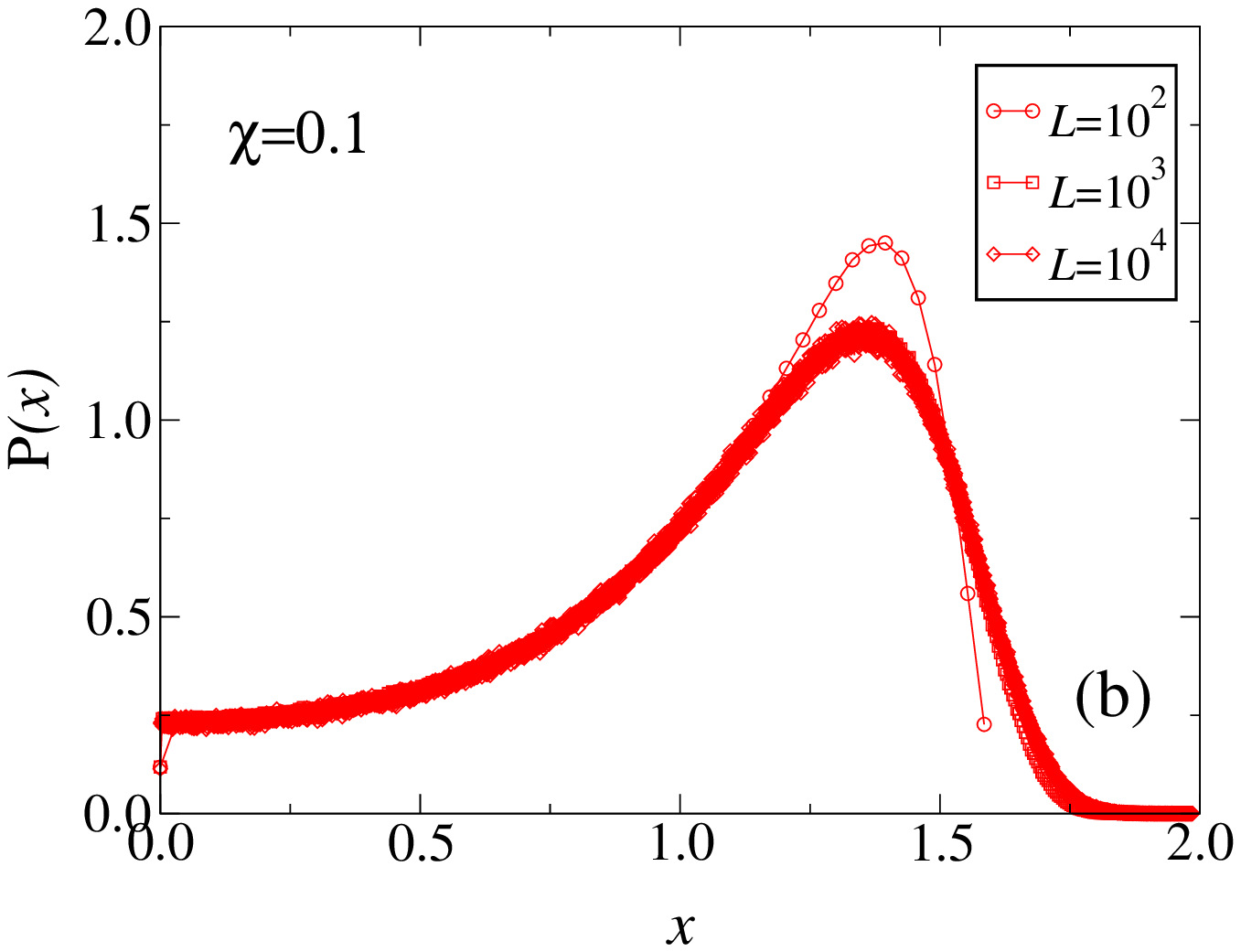}
\includegraphics[width=0.45\linewidth]{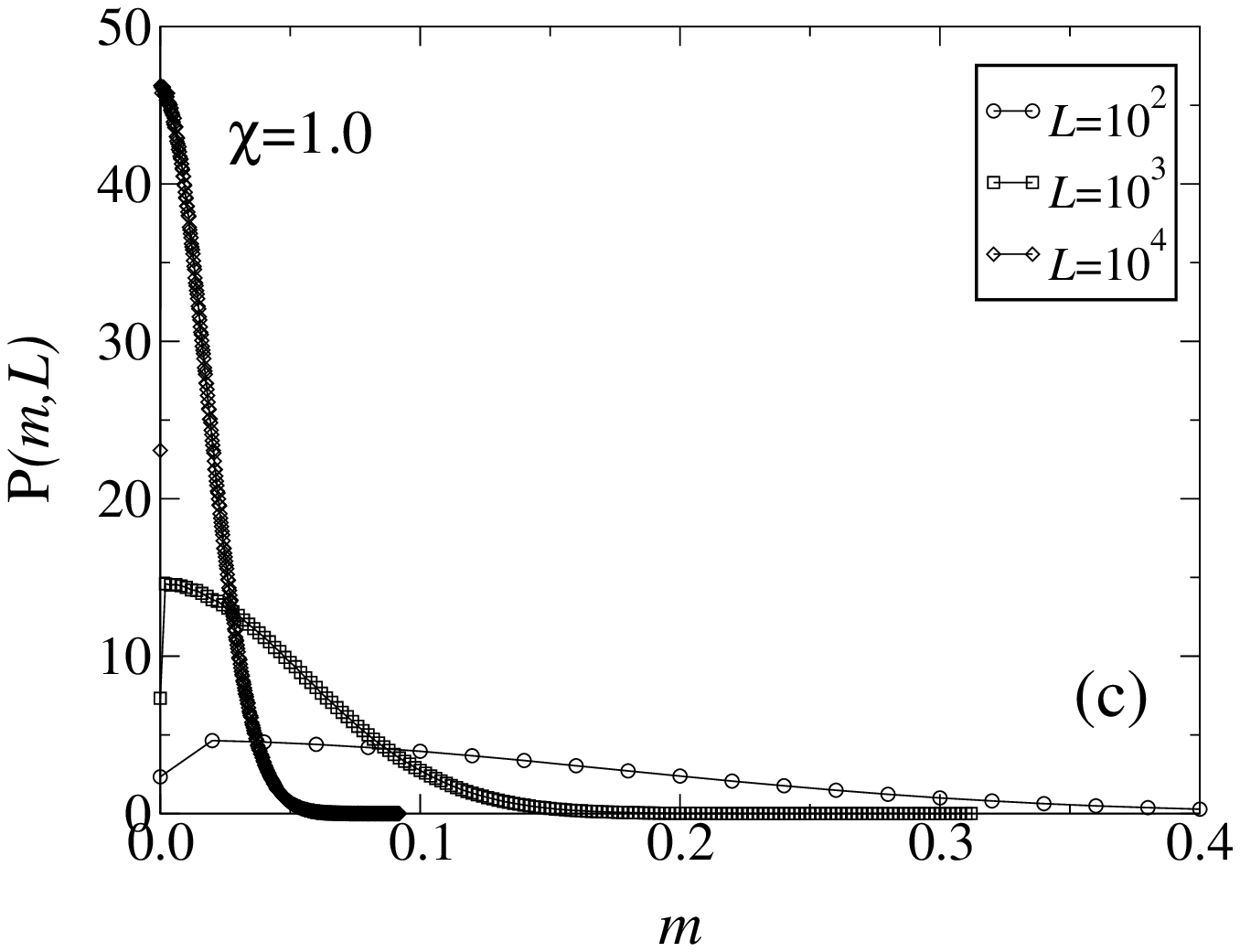}
\includegraphics[width=0.45\linewidth]{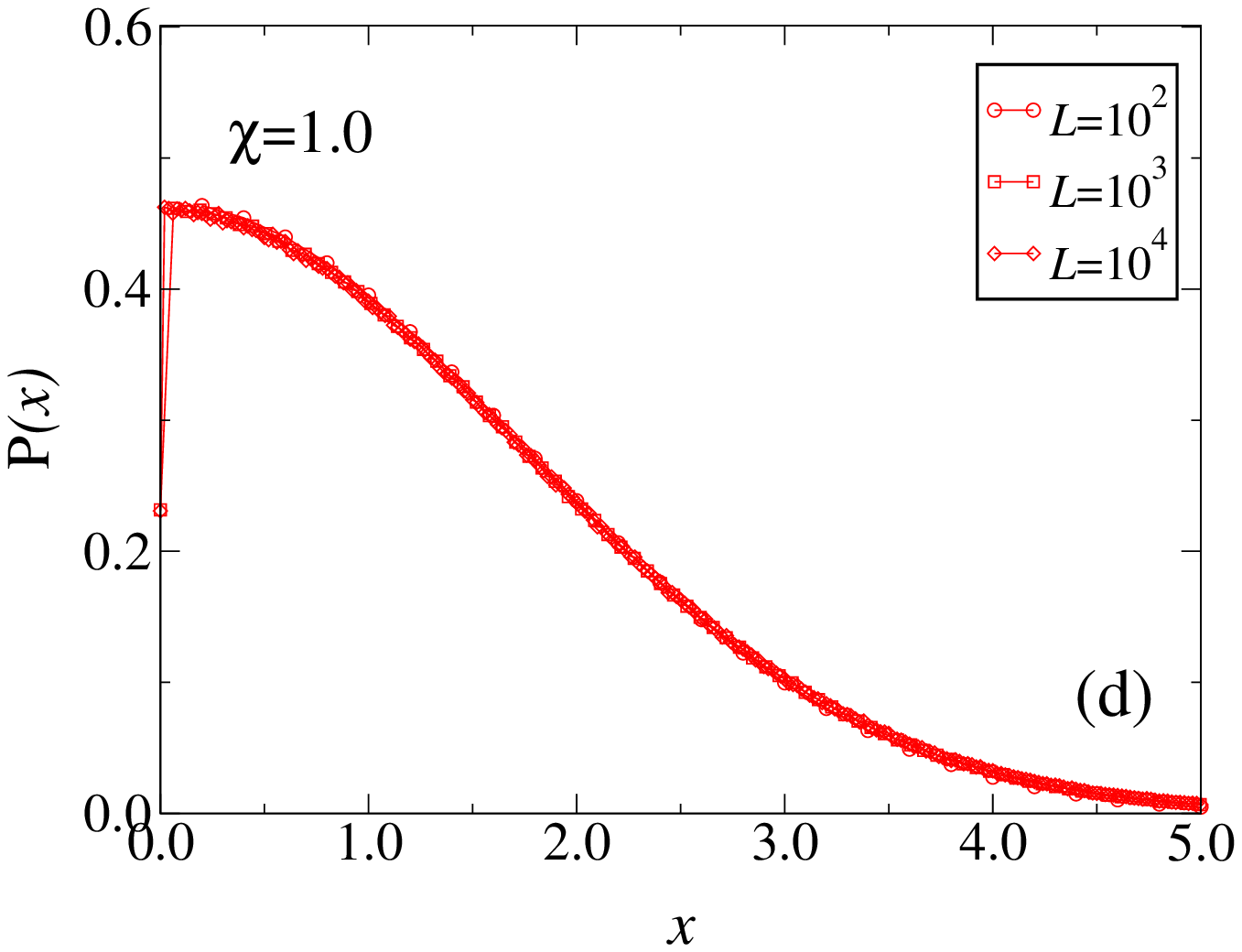}
\caption{\small Numerically measured distribution functions $P(m,L)$ and the corresponding scaling functions $\mathcal{P}(x)$ for $L=10^2,10^3,10^4$. (a,b) Regime (i) with $\chi=0.1$ and $\mathcal{P}(x)=L^{-0.1}P(L^{-0.1}x,L)$. (c,d) Regime (ii) with $\chi=1.0$ and the Gaussian scaling function $\mathcal{P}(x)=L^{-1/2}P(L^{-1/2}x,L)$. The deviations for $L=10^2$ for both regimes are due to the fact that the analytic asymptotics is only valid for large $L$. The ensemble averages were obtained by averaging over $10^7$ sequence realizations for each parameter setting with random initial conditions, resulting in symmetric distributions (only positive values shown).
\label{mkl_distribution}}
\end{figure}

We limit ourselves here to evaluating the equilibrium distribution  $P(m,L)$ 
asymptotically for large values of $L$. The solution of~(\ref{dynamics_Pm}) 
defines different parameter regimes:

\begin{enumerate}
\item{Strong correlation regime ($\chi<1/2$):} The large-$L$ asymptotics is determined by balancing dilatation and deterministic decay, i.e., the first two terms on the r.h.s.~of equation~(\ref{dynamics_Pm}). For this regime, we obtain 
\begin{equation}
P(m,L)=L^{\chi}\mathcal{P}_\chi (x) \quad 
\mbox{with}\quad x = mL^\chi,
\label{P1}
\end{equation}
where $\mathcal{P}_\chi (x)$ is a scaling function (whose form is determined by the stochastic dynamics
on smaller scales). We can verify the consistency of the solution~(\ref{P1}) by checking
that the third term on the r.h.s.~of~(\ref{dynamics_Pm}) gives a contribution which is 
subleading by a factor $L^{2 \chi -1}$ for large $L$.  This result is also verified by our numerics, as shown in figures~\ref{mkl_distribution}(a,b), where we present measured distributions $P(m,L)$ and the collapse into one scaling function $\mathcal{P}_\chi (x)$. Obviously, the scaling of $P(m,L)$ also determines the scaling of its moments 
$\langle m^k\rangle (L) \equiv \int m^k P(m,L)dm$ ,
\begin{equation}
\label{msym}
\langle m^k\rangle (L) \propto L^{-k\chi}.
\end{equation}  
This is consistent with the scaling of the one-point and two-point functions, obtained in equations~(\ref{s_k_asymptotics}) and~(\ref{C_asymptotics}).

\item{Weak-correlation regime ($\chi>1/2$):} Equation~(\ref{dynamics_Pm}) 
has an exact solution of Gaussian form, 
\begin{equation}
\fl\qquad P(m,L)=\sqrt{\frac{L}{2\pi \xi(\chi)}}
\exp \left [-\frac{(m - m_0 L^{-\chi})^2 L}{2\xi(\chi)} \right ]
\quad\mbox{with}\quad\xi(\chi) = \frac{\chi}{\chi - 1/2}. 
\label{P2}
\end{equation}
This solution has the expectation value 
\begin{equation}
\langle m \rangle (L) = m_0 L^{-\chi}
\label{ave}
\end{equation}
(with the coefficient $m_0$ determined by the initial condition) and
the variance 
\begin{equation}
\langle m^2 \rangle (L) - \langle m \rangle^2 (L) = \frac{\xi(\chi)}{L}.
\label{var}
\end{equation}
It is thus of similar form as the simple fluctuation-dissipation
equilibrium $\exp[-m^2/2L]$ for $\lambda = 0$, obtained from the last two terms
on the r.h.s.~of~(\ref{dynamics_Pm}). The transport term generates an
additional length scale $\xi$ since individual  sites are not completely independent
of each  other but are strongly correlated on scales smaller than $\xi$ due to
duplications. This reduces the number of effectively independent fluctuating
sequence segments to $L/\xi$. Numerical measurements of the distribution
$P(m,L)$ in this regime for random initial conditions ($ m_0 = 0$) 
and the corresponding scaling function 
$\mathcal{P}_\chi(x) \propto \exp[-x^2/2 \xi(\chi)]$ with $x \equiv mL^{1/2}$
are shown in figures~\ref{mkl_distribution}(c,d).  

\item{Transition point ($\chi = 1/2$):} The solution of (\ref{dynamics_Pm}) 
is still of Gaussian form,
\begin{equation}
P(m,L)=\sqrt{\frac{L}{2\pi \log L}}
\exp \left [ -\frac{(m - m_0 L^{-\chi})^2 L}{2 \log L} \right ]. 
\end{equation}
\end{enumerate}

The existence of two different scaling regimes has direct consequences for the detectability 
of correlations from data of a single sequence on large scales. In the strong-correlation regime
($\chi < 1/2$), the composition bias on arbitrary large scales $L$ is determined primarily by 
the ancestral bias, while the mutational fluctuations can be neglected asymptotically. In the 
weak-correlation regime, the ancestral bias can only be detected on scales $L < L^*$, while 
the mutational noise is dominant on larger scales. The scale $L^*$ can be estimated by equating the 
average $\langle m\rangle (L^*)$ with the rms. deviation 
$(\langle m - \langle m \rangle)^2 \rangle (L^*))^{1/2}$ 
given by equations~(\ref{ave}) and (\ref{var}). 

The difference between the strong- and weak-correlation regime is illustrated in figure~\ref{two-regimes}, where we show
two single sequences generated from an ancestor letter $+1$. In the strong-correlation regime,
the entire sequence has a detectable bias towards $+1$, with islands of $-1$ tracing back to 
their ancestors generated by mutation events (figure~\ref{two-regimes}(a)). In the weak-correlation regime, the
sequence is seen to consist of strongly correlated segments of length  
$\xi \approx 5$, but it looks random on larger scales (figure~\ref{two-regimes}(b)). 

We stress again that the existence of two different scaling regimes with a
transition at $\chi=1/2$ is a feature of the full distribution $P(m,L)$ in the
asymptotic regime $L \gg 1$. Expectation values such as the composition
bias~(\ref{s_k_asymptotics}) and the correlation function~(\ref{C_asymptotics})
have a universal form in both regimes and no transition at $\chi = 1/2$.

\begin{figure} [t!]
\begin{itemize}
\item[(a)]{\tiny$+++++++++++++++++++++++++++++++++++++++++++++++++++++++++-+-+++++++-+-+-----+++++++-++--++++++-----++----------------------------+++++++---------------+--++++++++++++++++++---+++++++++-++++++--+++-++++++++++++++++++++--++++++++++++++++++++---+++++++++++++--+++++++++++++++++++++++++++++++++++++++++++++++++++++++++++--+-+++++++++++++++++++++++++---+-++++++++++++++++++--++++++++++++++++++++++++++++++$}
\item[(b)]{\tiny$++-----++-+-++--+++---++++---++++++++--+++++-++++---+-----+-+-----++-++-------++++------++------------+-+++++-++++----+++++++++-+++-+++++++++--++------++--+----++-+++++++++-+-----++-+--+--+------++--++--+--+----++----++++++++---+-+-+--++--+++-++-++++--+-----+++------+-+++-+++++-----++++++-+--++-++++++---+-+--+++++--+-+--+-----+--+-+++--++++++---++++++++--+++----+++-----++-----+++++-+--++++--++++-+$}
\end{itemize}
\caption{\small A single sequence of length $N=400$ generated by the expansion-randomization process from
an initial letter $+1$.
(a) Strong-correlation regime ($\mu=0.5$, $\delta_1=10.0$, i.e.~$\chi=0.1<1/2)$: The sequence retains a net composition bias 
towards $+1$ in its entire length, i.e., the initial composition bias is detectable.
Minority islands of $-1$ are found on all scales. 
(b) Weak-correlation regime ($\mu=0.5$, $\delta_1=1.0$, i.e.~$\chi=1.0>1/2$): The sequence consists of strongly correlated 
islands of length $\xi \approx 5$ but looks random on larger scales. The initial composition
bias is not detectable.\label{two-regimes}}
\end{figure}

\section{Model extensions and symmetry breaking}

\label{model_extensions}

\subsection{Biased insertions}

In the following, we will investigate a generalization of the dynamical model and
thereby demonstrate the universality of our approach. For simplicity, we start
with a single-letter model ($\ell_{\rm max}=1$). In contrast to the original
model of section~\ref{definition}, where random insertions were defined as the
insertion of random letters $x=\pm 1$ at position $k+1$, which was independent
of the preceding sequence element $s_k$, we now want to consider biased
insertions. This extension is biologically well motivated, since there is ample
evidence by now that the rates of segmental insertions into the genome, as
e.g.~those of interspersed repeats, are biased by the local GC-content
of the genomic region~\cite{Lander01}. Formally, the biased insertion process in
our model is defined by      
\begin{equation}
(\cdots,s,\cdots)\quad\to\quad(\cdots,s,y[s],\cdots)\qquad\mbox{insertion
rate}\;\eta,
\end{equation}
where $y[s]$ denotes a randomly chosen letters $y[s]=\pm 1$ with an average bias
depending on the value of the preceding sequence element $s$,
\begin{equation}
\langle y[s]\rangle=\nu s,\qquad\nu\in[-1,1].
\end{equation}
The degree of dependence can thereby be tuned by a parameter $\nu$. In fact, the
random insertions of the original model are the special case of this generalized
process using $\nu=0$, while $\nu=1$ corresponds to duplications.

The contributions of this process to the dynamics of the joint-probabilities
$P_{\rm eq/op}(r,t)$ can still be calculated exactly. (\ref{master_a}) and
(\ref{master_e})-(\ref{master_h}) will not be affected, since the biased
insertion process will neither change the effect of single-site mutations, nor
the ``shift'' and ``transport'' of joint-probability. However, an additional
multiplicative factor $(1-\nu)$ has to be incorporated in~(\ref{master_b})
and~(\ref{master_c}), while effects on~(\ref{master_d}) are described by an
additional factor $(1+\nu)$. Concerning the Master equation for $C(r)$ in the
continuum limit~(\ref{C_continuum}), this biased insertion process does
therefore not affect the asymptotic growth rate $\lambda$, but the effective
mutation rate is now given by
\begin{equation}
\mu_{\rm eff}=\mu+\frac{1}{2}(1-\nu)\eta.
\end{equation}    
We shortly want to mention that the biased insertion of single letters can
generically be extended to the biased insertion of segments $(y[s])_{\ell}$ at a
rate $\eta_{\ell}$ with an average bias of their elements $\langle y_i[s]
\rangle=\nu_{\ell}s$. In this case, one might actually have
$\nu_{\ell}=\nu(\ell)$, and asymptotically for the effective mutation rate we
yield
\begin{equation}
\mu_{\rm eff}=\mu+\frac{1}{2}\sum_{\ell=1}^{\ell_{\rm
max}}(1-\nu_{\ell})\eta_{\ell}.
\end{equation}

\subsection{Biased mutations and symmetry breaking}

\label{symmetry_breaking}

The model considered so far was symmetric concerning $s_k\to -s_k$, i.e., the
rates of all processes were independent of $s_k$. However, it is known that this
symmetry is not granted for genomic evolution. For example, distinct mutation
rates of different nucleotides lead to the unequal frequencies of the four
different nucleotides along genomic DNA~\cite{Arndt03}. In the following we will
show that the restriction to symmetric processes is not crucial concerning the
emergence of long-range correlations and the universal scaling of the generated sequences. A simple
scenario breaking the model's $Z_2$ symmetry is the choice of asymmetric
mutation rates,
\numparts
\begin{eqnarray}
(\cdots,+1,\cdots)&\quad\to\quad(\cdots,-1,\cdots)&\qquad\mbox{rate}\;\mu^+\\
(\cdots,-1,\cdots)&\quad\to\quad(\cdots,+1,\cdots)&\qquad\mbox{rate}\;\mu^-,
\end{eqnarray}
\endnumparts
with $\mu^+\neq \mu^-$. In this case, the Master equations of the probabilities
$P^{\pm}_k(t)$ are
\begin{eqnarray}
\frac{\partial}{\partial
t}P^{\pm}_k(t)=&\pm\mu^-P^{\mp}_k\mp\mu^+P^{\pm}_k+\sum_{\ell=1}^{\ell_{\rm
max}}\mbox{min}(k-1,\ell)\:\gamma_{\ell}^+\:(1/2-P_k^{\pm})\nonumber\\
&+O\left(\sum_{\ell=-\ell_{\rm max}}^{\ell_{\rm
max}}P^{\pm}_{k+\ell}-P^{\pm}_k\right),
\end{eqnarray} 
and we have already shown in section~\ref{average_bias} that asymptotically
$P^{\pm}_k$ is independent of $k$ if all sequence sites $s_k$ are allowed to
mutate. Thus, for the asymptotic stationary average composition bias $\langle
s_k\rangle=P^+-P^-$ in the asymmetric mutation model we obtain
\begin{equation}
\label{s_k_assymetric}
\langle s_k \rangle=\frac{\mu^--\mu^+}{\mu^-+\mu^++2\gamma^+_{\rm eff}}.
\end{equation}    
Concerning the dynamics of the joint probabilities $P_{\rm eq/op}(r,t)$, the
introduction of asymmetric mutation rates will only change the mutational term,
while the contributions of duplications, random insertions, and deletions will
not be affected. In the asymmetric model, the Master equations for $P_{\rm
eq/op}(r,t)$ are now given by
\numparts
\begin{eqnarray}
\fl\qquad\quad\frac{\partial}{\partial t}P_{\rm eq}(r,t)=&+(\mu^++\mu^-)P_{\rm
op}(r)-2\mu^+P^{++}(r)-2\mu^-P^{--}(r)+Q_{\rm eq}(r,t)\\
\fl\qquad\quad\frac{\partial}{\partial t}P_{\rm op}(r,t)=&-(\mu^++\mu^-)P_{\rm
op}(r)+2\mu^+P^{++}(r)+2\mu^-P^{--}(r)+Q_{\rm op}(r,t),
\end{eqnarray}  
\endnumparts
where $P^{++/--}(r)$ are the joint probabilities of simultaneously finding
$s_k=s_{k+r}=+1$ and $s_k=s_{k+r}=-1$, respectively. $Q_{\rm eq}(r,t)$ denotes
the terms~(\ref{master_b})-(\ref{master_h}) with the $k$-dependence of $P_{\rm
eq/op}(r,t)$ already dropped, while $Q_{\rm op}(r,t)$ is obtained by exchanging
$P_{\rm eq}$ and $P_{\rm op}$. The dynamics of $C(r,t)$ in the asymmetric model
is therefore 
\begin{equation}
\fl\qquad\quad\frac{\partial}{\partial t}C(r,t)=-2(\mu^++\mu^-+\gamma^+_{\rm
eff})\:[C(r)+\langle s_k \rangle^2]+[Q_{\rm eq}(r,t)-Q_{\rm op}(r,t)],
\end{equation} 
where we used~(\ref{s_k_assymetric}) and $\langle
s_k\rangle=P^+-P^-=P^{++}(r)+P^{+-}(r)-P^{-+}(r)-P^{--}(r)$ with
$P^{+-}(r)=P^{-+}(r)$. 
Defining the effective mutation rate of
the asymmetric model,
\begin{equation}
\tilde{\mu}_{\rm eff}=\frac{1}{2}(\mu^++\mu^-+\gamma^+_{\rm eff}),
\end{equation}
the stationary solution of this dynamics in the continuum limit is now given by
\begin{equation}
\label{casym}
C(r)\propto r^{-\alpha}+ \langle s_k\rangle^2\quad\mbox{with}\quad\alpha=2\chi=\frac{4\tilde{\mu}_{\rm eff}}{\lambda}.
\end{equation}
The magnitude of the segmental composition bias~(\ref{m}) scales as
\begin{equation}
\label{masym}
\langle |m(L)|\rangle \propto L^{-\chi}+ \langle s_k\rangle.
\end{equation}
Hence, breaking the $Z_2$ symmetry by introducing asymmetric mutation rates will
not change the long-range correlations and the general scaling of the model. It is obvious from
equations~(\ref{casym}) and~(\ref{masym}) that the scaling still holds for the
connected correlation function $C^c(r)\equiv\langle s_k s_{k+r}\rangle-\langle
s_k \rangle^2$ and the shifted segmental composition bias $\langle1/L|\sum_{k'=k}^{k+L-1}s_{k'}|\rangle-\langle s_k\rangle$. 

\subsection{Universality}

\label{universality}

The structure of equation~(\ref{C_continuum}) reveals the basic mechanisms
generating long-range correlations in a very general class of
expansion-randomization systems that share three fundamental characteristics of
their dynamics. The first feature is an overall exponential expansion of the
system transporting correlations from shorter to larger sequence distances
(combined effects of duplications, insertions, and deletions in our model).
Mathematically, this transport is described by a dilatation operator
$r\partial/\partial r$ (second term on the r.h.s.~of~(\ref{C_continuum})). On
the other hand, all correlations are counteracted by local processes randomizing
the sequence (mutations) and therefore trying to diminish $C(r)$ (first term
of~(\ref{C_continuum})). The competition between expansion and randomization
results in an algebraically decaying $C(r)\propto r^{-\alpha}$ in the stationary
state, with $\alpha$ determined by a simple ration of effective growth rate and
effective mutation rate. Calculation of these two fundamental parameters for any
set of processes constituting such system determines the large-distance
asymptotics of the correlations in the generated sequences. However, $C(r)=0$
for all $r$, is also a stationary solution of equation~(\ref{C_continuum}).
Hence, in order for long-range correlations to be established, a third necessary
feature of such systems is the presence of a mechanism continuously producing
correlations on short scales. They serve as an ongoing reservoir for the
transport of correlations to larger sequence distances and ensure the existence
of a non-zero value $C(r_0)>0$ for a specific $r_0\ge 1$ (in our model, these
initial correlations on short-scales are produced by duplications). As an
intuitive example for the necessity of this third condition, consider an
expansion-randomization system with mutations and insertions of single random
letters, but no duplications. This system features exponential expansion, as
well as local randomization. But the insertion process is not capable of
producing $C(1)>0$, and therefore no long-range correlations can be established
in the generated sequences. 

As expected from standard scaling theory, the decay  of the two-point function has twice the exponent as the corresponding decay of the one-point 
function. The value $\chi$ can be interpreted as the scaling 
dimension of the variable $s_k$ in this universality class. There is a 
1-parameter family of decay exponents as, for example, in the  
Gaussian model in two dimensions. This universal behavior is unaffected by 
the breakdown of the $Z_2$ symmetry, which manifests itself only in the 
non-universal constants in~(\ref{masym}) and~(\ref{casym}).

\subsection{Numerical Analysis}

\label{numerical_implementation}

Numerical simulation of the stochastic sequence
dynamics~(\ref{model_definition}) was implemented using a Monte Carlo
procedure. During each discrete time step 
\begin{equation}
\Delta
t=\epsilon\cdot[(\mu+\sum\nolimits_{\ell}[\delta_{\ell}+\gamma_{\ell}^++\gamma_{\ell}^-
])N(t)]^{-1}
\end{equation}
with a tunable parameter $\epsilon\le1$, we choose a random site and randomly let a
process act on it. The probability $p_{\alpha}$ of a process $\alpha$ being
executed on the drawn site is
\begin{equation}
p_{\alpha}={\rm rate}(\alpha)\cdot\Delta t.
\end{equation}
The overall probability of executing any process on the drawn site therefore
depends on the parameter $\epsilon$. While $\epsilon=1$ assures exactly one process being
executed, for small $\epsilon$, on he other hand, no process will be chosen to act on
the drawn sites in most of the cases. We use $\epsilon=0.1$ for our numerical
simulations. 

For a single realization of the stochastic dynamics,
the average segmental composition bias $\langle|m(L)|\rangle$ and the 
correlation function $C(r)$ are well approximated by sequence averages,
\begin{eqnarray}
\label{C_seq}
\langle|m|\rangle (L) &\approx&\frac{1}{N-L}\sum_{k=1}^{N-L}\frac{1}{L}
\left|\sum_{k'=k}^{k+L-1} s_{k'}\right|, 
\\
C(r)&\approx&\frac{1}{N-r}\sum_{k=1}^{N-r} s_k s_{k+r}, 
\end{eqnarray}
for sufficiently small values of $r$ and $L$ to allow efficient averaging. 
Averaging over 100 sequence realizations reduces the noise further and 
produces very accurate measurements of $\langle|m|\rangle(L)$ and $C(r)$.

If the dynamics obeys $Z_2$ symmetry, we can directly infer the decay 
exponent $\alpha$ from these measurements, according to equations~(\ref{msym})
and (\ref{csym}). However, if the $Z_2$ symmetry is violated, these power
laws have to be disentangled from the additional constants $\langle s_k
\rangle$ respectively $\langle s_k\rangle^2$, see equations~(\ref{masym}) and
(\ref{casym}). If the microscopic processes are known, these non-universal 
constants can be calculated. A numerical problem arises, however, in the analysis
of genomic DNA sequences, where the $Z_2$ symmetry is broken by an 
unknown amount. In that case, we can self-consistently fit the data in the
form $\langle|m|\rangle(L) = a L^{-\chi} + c$ and 
$C(r) = b r^{-2 \chi} + c^2$. Hence, the link between the finite-size
scaling of $\langle|m|\rangle(L)$ and the scaling of the correlation function $C(r)$ dictated by 
universality is of practical importance for data analysis. In particular, it is not justified in general to approximate the constant $c$ by $1/N\sum_{k=1}^Ns_k$ for sequences of finite length $N$ in the strong correlation regime $\chi<1/2$, as it is often done in the literature. Furthermore, we can check consistency
with the exponent $\beta = 1 -2\chi$ of the GC power spectrum. 
Power spectra can easily be obtained using
the \emph{Fast Fourier Transform} algorithm~\cite{Press97}.

\section{Dynamical correlations}

\label{dynamical_correlations}

\subsection{Correlation build-up}

\label{built_up}

\begin{figure} [t!]
\centering
\includegraphics[width=0.49\linewidth]{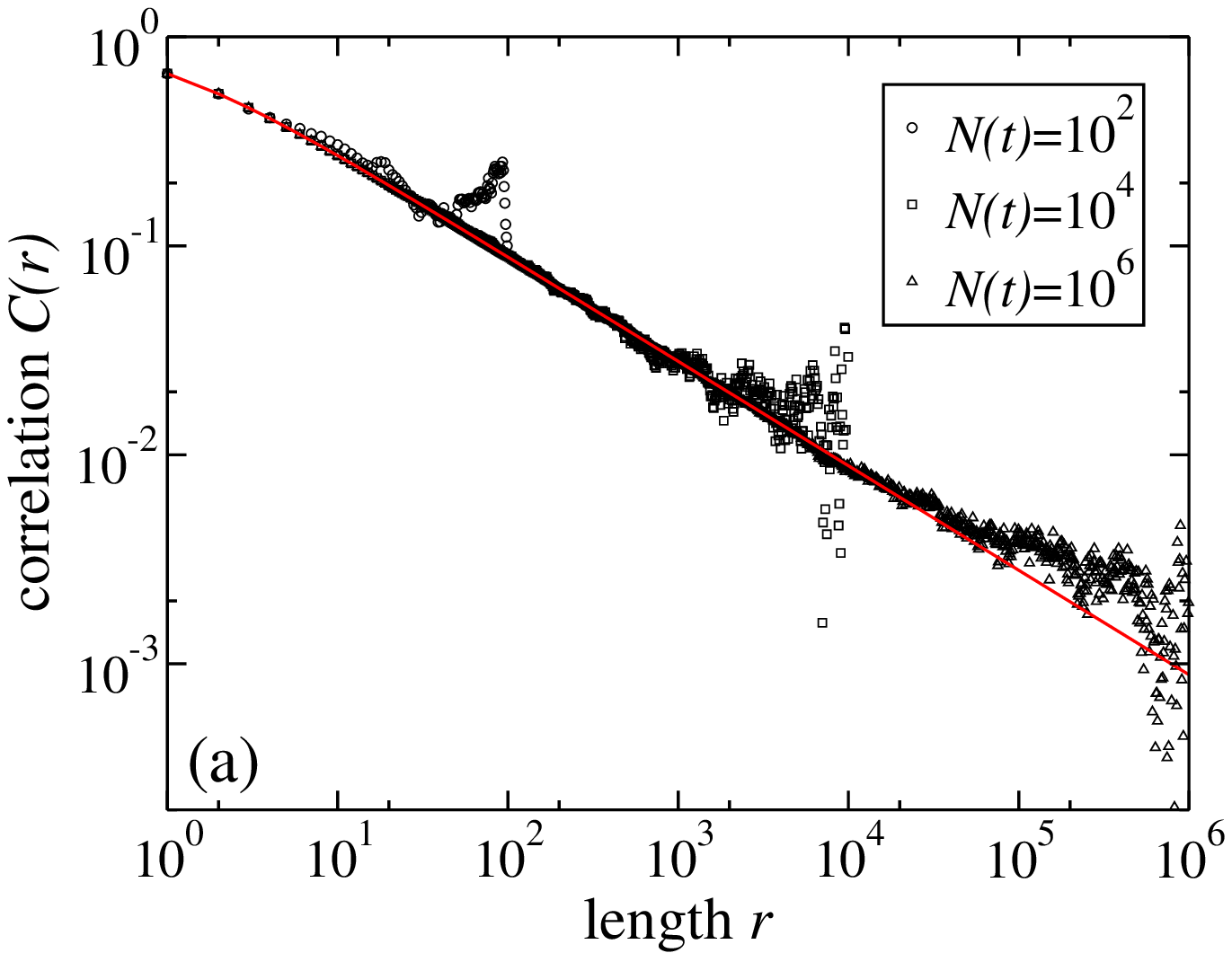}
\includegraphics[width=0.49\linewidth]{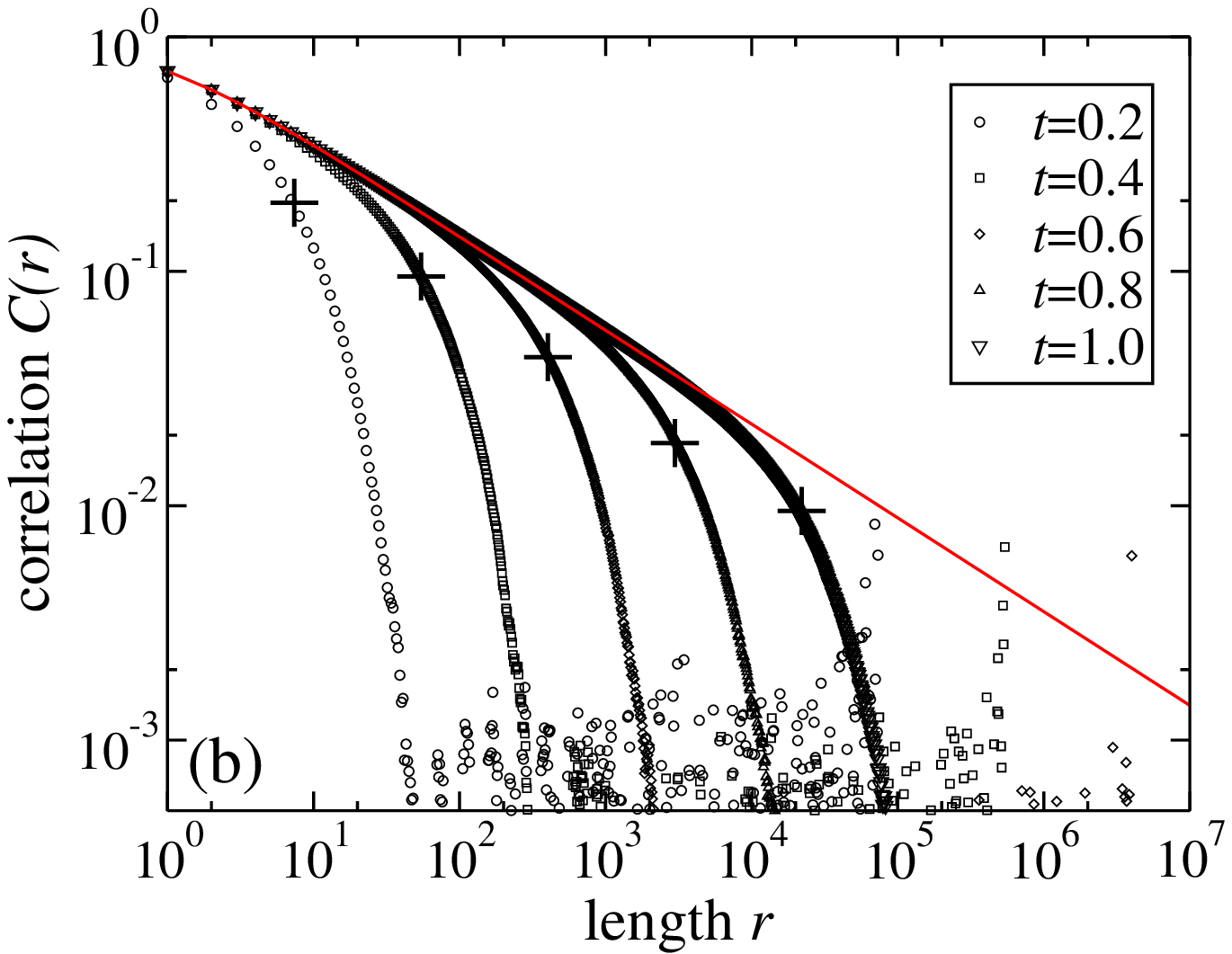}
\caption{\small Time-dependent correlations $C(r,t)$. (a) Build-up of long-range
correlations by stationary growth. Measured $C(r,t)$ at various intermediate
lengths $N(t)=10^2,10^4,10^6$ (symbols) together with the stationary
form~(\ref{C_dup_mut_result}) for $\mu=1.0$, $\delta_1=8.0$ (line). (b)
Correlation build-up from a random sequence of length $N_0=10^4$. At $t=0$ the
processes started acting on the sequence with rates $\mu=1.0$, $\delta_1=10.0$.
Measured $C(r,t)$ (symbols) of the simulated sequences after various times $t$
(averages over 100 realizations). Black crosses denote the corresponding mean
sizes $r^*(t)=\exp(\lambda t)$. Correlations have been established in the
sequences according to their analytic stationary form (red line) in the regime
$r<r^*(t)$, while they vanish for $r>r^*(t)$.   
\label{C_growth_plot}}
\end{figure}

Up to now, results for the correlations $C(r)$ in our model have only been
obtained for the stationary state, reached in the limit
$t\rightarrow\infty$. We now  take a closer look on the dynamical aspects
of the build-up of correlations in growing sequences. Starting with a sequence
$S(t=0)=(x)$, where $x=\pm 1$ denotes a uniformly distributed random letter, the
correlations are found to be present from the beginning.
Figure~\ref{C_growth_plot}(a) gives examples for $C(r)$ measured along short
single sequence realizations of length $N(t)=10^2$, $10^4$, and $10^6$.

But of course, correlations cannot be present right from the beginning on all
scales if we use a sequence $S(t=0)=(s_1,...,s_{N_0})$ with length $N_0>1$ as
initial condition, whose letters are randomly chosen (and thus uncorrelated).
All the processes of our model are local processes: a single step can introduce correlations only up to a microscopic length-scale $\ell_{\rm
max}$. Thus, there will be a cutoff-length $r^*(t)$, up to which correlations
can have been established at time $t>0$. It is determined by the average
distance, two copies of a duplication event at $t=0$ are separated from each
other along the sequence at time $t$. Therefore we have 
\begin{equation}
\label{lightcone_size}
r^*(t)=\ell_{\rm max}\exp{(\lambda t)}.
\end{equation}
Fig~\ref{C_growth_plot}(b) shows that $r^*(t)$ marks the range where $C(r)$ will
start to deviate significantly from its stationary form.

\subsection{Distinct dynamical regimes and correlation decay}

\begin{figure} [t!]
\centering
\includegraphics[width=0.49\linewidth]{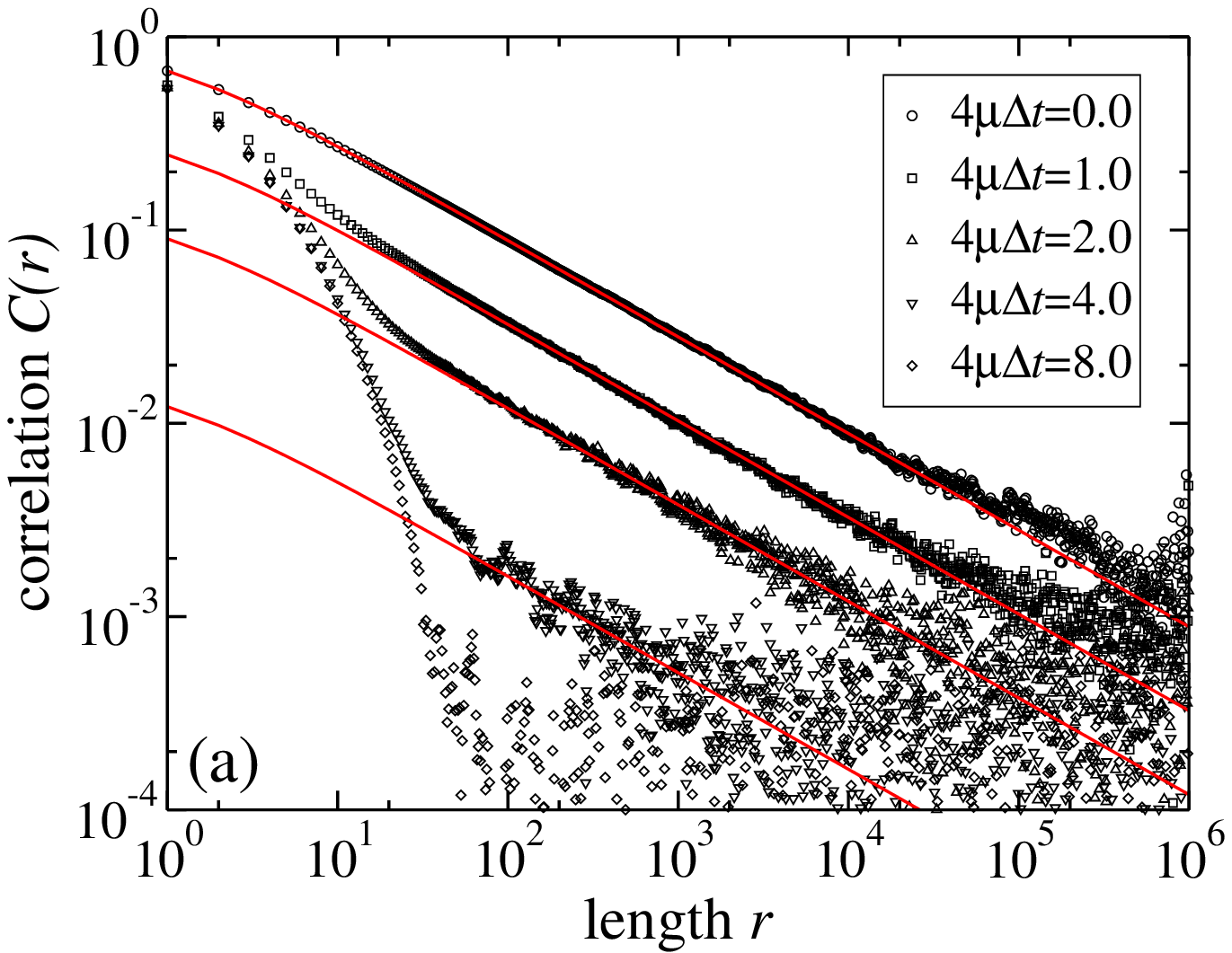}
\includegraphics[width=0.49\linewidth]{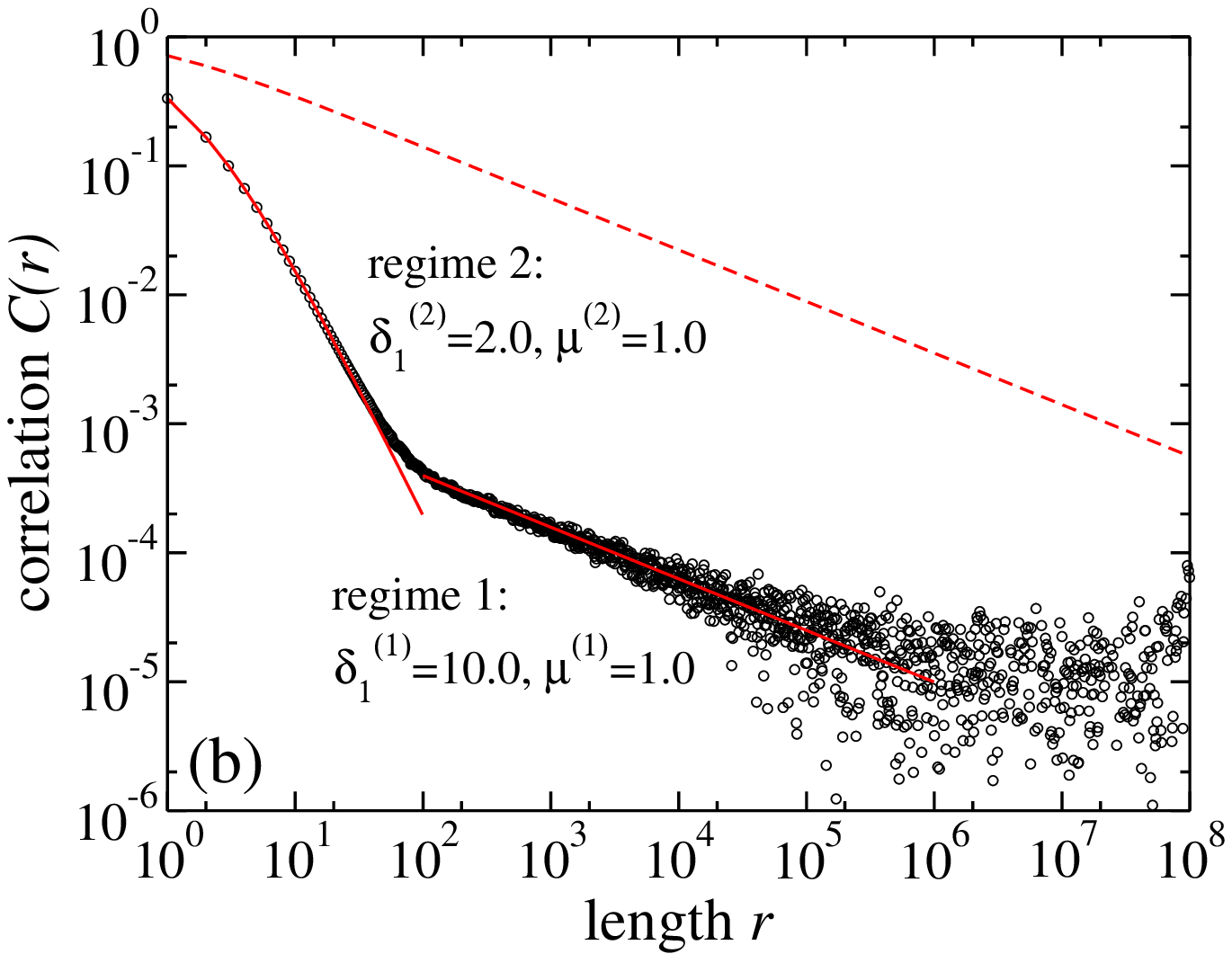}
\caption{\small (a) Decay of correlations during sequence evolution at
stationary length $N_0=10^6$. Measured $C(r,t)$ at various times $\Delta t$
(symbols) together with the analytic decay of the long-range tail given by
equation~(\ref{C_decay}). In the previous growth phase for $t<t_0$, correlations
have been established by a single-letter duplication-mutation dynamics with
$\mu=1.0$ and $\delta_1=8.0$ until the sequences reached the length $N_0=10^6$.
For $\Delta t =t-t_1>0$, a single-letter deletion process with $\gamma^-_1=8.0$
was introduced. Note that the correlations on short scales are preserved during
the second phase. (b) $C(r)$ with two scaling regimes 1 and 2 (symbols). Process
rates are: $\mu^{(1)}=1.0,\:\delta_1^{(1)}=10.0$ and
$\mu^{(2)}=1.0,\:\delta_1^{(2)}=2.0$. The dashed red line is the analytical $C(r,t)$ for the parameters of phase 1. The second phase lasted over a
period of time that on average allowed the sequences to increase their length by
a factor of 100. For each scaling regime $(n=1,2)$, $C(r)$ obeys the predicted
algebraic decay with exponent $\alpha^{(n)}=4\mu_{\rm eff}^{(n)}/\lambda^{(n)}$.
The transition between both regimes is sharp and its position agrees with the
value predicted by~(\ref{lightcone_size}).\label{C_decay_plot}}
\end{figure}
There is ample evidence that the rates of local evolutionary processes are not
constant in time~\cite{Arndt03}. We mimic this non-stationarity of the individual process rates
by the succession of several distinct dynamical phases. For each individual
phase $n$, the rates of the elementary processes are constant during the time
interval $t_{n-1}<t<t_n$ and result in specific values of $\lambda^{(n)}$ and
$\mu_{\rm eff}^{(n)}$ for that particular phase. Between different phases,
however, the complete set of rates may change, 
\begin{equation}
\begin{array}{lll}
\mbox{phase
1:}\quad&(\mu^{(1)},\:\delta_1^{(1)},\:\cdots\:)\quad&\mbox{for}\quad
t_0<t<t_1\\
\mbox{phase
2:}\quad&(\mu^{(2)},\:\delta_1^{(2)},\:\cdots\:)\quad&\mbox{for}\quad
t_1<t<t_2\\
:&:&:\\
\mbox{phase
$n$:}\quad&(\mu^{(n)},\:\delta_1^{(n)},\:\cdots\:)\quad&\mbox{for}\quad
t_{n-1}<t<t_n\\
:&:&:
\end{array}
\end{equation}
Using the findings of section~\ref{built_up}, we can generalize our dynamics
with respect to varying rates during sequence evolution. We start with the
following simple two-stage scenario: sequence growth with rate $\lambda^{(1)} >
0$ for $0<t<t_1$, followed by a second phase with $\lambda^{(2)} = 0$ and
therefore $\langle N \rangle(t) = N^{(1)}$ for $t > t_1$. It is obvious from
equation~(\ref{C_continuum}) that stationary long-range correlations only emerge
as long as the sequence grows, i.e.~for $\lambda^{(n)} > 0$.  The time-dependent
solution of~(\ref{C_continuum}) for the asymptotics of $C(r)$ during the second
phase $(t>t_1)$ then takes the form
\begin{equation}
\label{C_decay}
C(r,t) = C(r,t_1)\:e^{-4\mu_{\rm eff}^{(2)}\Delta t}\propto r^{-4 \mu_{\rm
eff}^{(1)}/\lambda^{(1)}}e^{-4\mu_{\rm eff}^{(2)}\Delta t}
\end{equation}
with $\Delta t=t-t_1$. Thus, the long-range tails of the correlations
established during the first phase are preserved in the second phase, but their
amplitude decays exponentially with a characteristic time scale $\tau=(4\mu_{\rm
eff}^{(2)})^{-1}$.

In the short range part, however, correlations may still be present depending on
the particular set of process rates chosen to assure $\lambda^{(2)}=0$. If, for
example, all rates $\delta_{\ell}^{(2)}$, $\gamma_{\ell}^{+(2)}$,
$\gamma_{\ell}^{-(2)}$ are zero in the second phase, the only process acting
will be mutation which exponentially destroys correlations uniformly along the
sequence, and thus the amplitude of $C(r)$ will decay according to
equation~(\ref{C_decay}) for all lengths $r$. The situation becomes more complex if
$\lambda^{(2)}=0$ is accomplished in the presence of duplications by a
compensatory increase of the deletion rate. In this case, the duplication
process will keep correlations present at short lengths since there is always a
finite probability that a site $s_k$ recently originated by a duplication of
$s_{k-1}$ (which again might be a duplication of $s_{k-2}$, and so on.) and was
not yet affected by a mutation event. Numerical results for this type of
two-phase dynamics are shown in figure~\ref{C_decay_plot}(a), verifying the
exponential decay of the long-range tail, predicted by equation~(\ref{C_decay}).

In a general evolutionary scenario, with several distinct dynamical phases and
arbitrary values of $\lambda^{(n)}$ and $\mu_{\rm eff}^{(n)}$ for each
particular phase, the functional characteristics of the correlations in the
generated sequences will be shaped by a combination of correlation build-up and
decay, according to the mechanisms which have been revealed above. During phase
$n$ with $\lambda^{(n)}>0$, correlations will be established with
$\alpha^{(n)}=4\mu_{\rm eff}^{(n)}/\lambda^{(n)}$, and they will approximately
range over a length scale $r=1,\dots,r_{\rm max}$ with $r_{\rm
max}=\exp(\lambda^{(n)}\Delta t_n)$. The correlations already present from the
previous phases will be transported to larger sequence distances. If they ranged
across an interval $r=1,\dots,N(t_{n-1})$ at the end of phase $n-1$, they will
be shifted to the interval $r=N(t_{n-1}),\dots,N(t_n)$ during phase $n$. The
long-range tails, however, will still obey the same exponent corresponding to
the effective rates of the original growth phase they have originated from.
Additionally though, they are at the mercy of mutations, and their amplitude
will therefore decay exponentially on all scales according to
equation~(\ref{C_decay}) with the effective mutation rate $\mu_{\rm eff}^{(n)}$.
A numerical example of a two-stage dynamics with two distinct scaling regimes is
shown in figure~\ref{C_decay_plot}(b).

Given the chronology of the process rates for all phases, we thus can in
principle predict the different scaling regimes of the correlation function.
Furthermore, given the measured $C(r)$ of a sequence generated under the
influence of our processes, we might be able to reconstruct the chronology of
the ratio of the effective rates $\lambda$ and $\mu_{\rm eff}$ back throughout
its evolutionary history. In practice, however, such an attempt will be confined
by two major constraints. At first, all of the above statements only apply to
the long-range tails of $C(r)$. Thus, in order to perspicuously identify the
decay exponent $\alpha$ of a certain rate regime, the net expansion during that
regime must have been sufficiently large. Moreover, since the correlations of
the previous phases decay exponentially with a time-scale $\tau=(4\mu_{\rm
eff})^{-1}$, the ratio $\lambda/\mu_{\rm eff}$ of the succeeding phases should
be high. Otherwise, previously established correlations will rapidly decay below
the fluctuation threshold $\Delta C=1/\sqrt{N(t)}$, and thus cannot be measured
any longer.

\section{Discussion}

In this article, we have investigated a broad class of
stochastic sequence evolution processes as 
possible causes of the observed
long-range correlations in genomic DNA sequences. The
emergence of such correlations is seen to be a robust 
feature of the entire class of models. They can be observed,
e.g., in the two-point function and in the finite-size distribution
of the composition bias. The power law behavior of these quantities is
linked by a dynamical scaling theory.

Clearly, further analysis of genomic data is needed to corroborate
or refute possible causes of the observed correlations.
Comparative genomics of closely related species is expected
to offer a more detailed view on the elementary evolutionary
processes shaping genomes. One has to keep in mind that genomic DNA is a
highly heterogeneous environment~\cite{Karlin93}: it consists of genes,
noncoding
regions, repetitive elements etc., and all of these functional substructures may
imprint their signature on the amount of correlations found in a particular
genomic region. If a local expansion-randomization dynamics proves indeed
responsible for these correlations, the universality
established in this paper is crucial for the biological
relevance. There is clearly a multitude of microscopic elementary
processes, whose individual rates may be small and
difficult to measure. These rates may vary
across sequences, between species and between phases of
evolutionary history. However, they enter the
composition correlations in the mesoscopic range -- for length scales
between $10^3$ and $10^6$ --  only via two effective
parameters, the effective growth rate and the effective
mutation rate. It is this fact that provides an explanation for
the ubiquity of long-range correlations and a way of testing
the theory in a quantitative way.
While the emergence of long-range
tails appears to be universal, the decay exponent is not.
This may also provide useful information on the expansion
history of genomes.

Biology has sometimes been characterized as a ``science of 
exceptions''. There is an amazing diversity of biological
species. Genomes encode that diversity, so the concept of
universality, which has proved so successful in physics,
would hardly seem to be applicable to biology at first
glance. However, this may well depend on the questions we 
ask, and even the above quote may have its exception.
Genomic correlations could be an example of universality 
in evolutionary biology.

\end{document}